\documentclass[aps,jmp,amsmath,amssymb,reprint]{revtex4-1}
\usepackage{graphicx}
\usepackage{dcolumn}
\usepackage{bm}
\usepackage{graphicx}
\usepackage{subfigure}
\usepackage{physics}
\usepackage[english]{babel}
\usepackage[usenames, dvipsnames]{color}
\usepackage{hyperref}
\usepackage[normalem]{ulem}
\usepackage[utf8]{inputenc}
\usepackage{romannum}
\usepackage{float}
\DeclareMathOperator{\erfc}{erfc}
\usepackage{bigints}
\usepackage{comment}
\begin{document}
\bibliographystyle{apsrev.bst}
\pagenumbering{arabic}
\renewcommand{\figurename}{FIG.}
\def\tablename{TABLE}

\title{Magneto-transport in the monolayer \texorpdfstring{$ \textrm{MoS}_\textrm{2}$}{} material system for high-performance field-effect transistor applications}










\author{Anup Kumar Mandia$^{a}$}
	\author{Rohit Kumar$^{b}$}	
 \author{Seung-Cheol Lee$^{c}$}
\author{Satadeep Bhattacharjee$^{a}$}
 \author{Bhaskaran Muralidharan$^{b,d}$}
  \thanks{corresponding author: bm@ee.iitb.ac.in}
  
\affiliation{$^{a}$Indo-Korea Science and Technology Center (IKST), Jakkur, Bengaluru 560065, India}
\affiliation{$^{b}$Department of Electrical Engineering, Indian Institute of Technology Bombay, Powai, Mumbai-400076, India}
\affiliation{$^{c}$Electronic Materials Research Center, KIST, Seoul 136-791, South Korea.}	
\affiliation{$^{d}$Centre of Excellence in Quantum Information, Computation, Science and Technology, Indian Institute of Technology Bombay, Powai, Mumbai-400076, India} 

\begin{abstract}
Electronic transport in monolayer MoS\textsubscript{2} is significantly constrained by several extrinsic factors despite showing good prospects as a transistor channel material. Our paper aims to unveil the underlying mechanisms of the electrical and magneto-transport in monolayer MoS\textsubscript{2}. In order to quantitatively interpret the magneto-transport behavior of monolayer MoS\textsubscript{2} on different substrate materials, identify the underlying bottlenecks, and provide guidelines for subsequent improvements, we present a deep analysis of the magneto-transport properties in the diffusive limit. Our calculations are performed on suspended monolayer MoS\textsubscript{2} and MoS\textsubscript{2} on different substrate materials taking into account remote impurity and the intrinsic and extrinsic phonon scattering mechanisms. We calculate the crucial transport parameters such as the Hall mobility, the conductivity tensor elements, the Hall factor, and the magnetoresistance over a wide range of temperatures, carrier concentrations, and magnetic fields. The Hall factor being a key quantity for calculating the carrier concentration and drift mobility, we show that for suspended monolayer MoS\textsubscript{2} at room temperature, the Hall factor value is around 1.43 for magnetic fields ranging from $0.001$ to $1$ Tesla, which deviates significantly from the usual value of unity. In contrast, the Hall factor for various substrates approaches the ideal value of unity and remains stable in response to the magnetic field and temperature. We also show that the MoS\textsubscript{2} over an Al\textsubscript{2}O\textsubscript{3} substrate is a good choice for the Hall effect detector. Moreover, the magnetoresistance increases with an increase in magnetic field strength for smaller magnetic fields before reaching saturation at higher magnetic fields. The presented theoretical model quantitatively captures the scaling of mobility and various magnetoresistance coefficients with temperature, carrier densities and magnetic fields. 
\end{abstract}
\maketitle
\section{Introduction}

\indent Due to their unique electrical, mechanical, structural, and optical properties, including ultimate body thickness, sizable and tunable bandgap, strong exciton binding, low dark current \cite{mak2016photonics}, and high room-temperature mobility, transition-metal dichalcogenides (TMDCs) \cite{manzeli20172d,kolobov2016two,chhowalla2015two} are a noteworthy class of the two-dimensional (2D) layered materials for next-generation semiconductor devices for nanoelectronic, optoelectronic and spintronic applications \cite{choi2017recent,mak2016photonics,yu2017analyzing,manzeli20172d,wang2012electronics,wilson1969transition,zibouche2014transition}.
TMDCs such as MoS\textsubscript{2}, WS\textsubscript{2}, MoSe\textsubscript{2}, and WSe\textsubscript{2} are the most promising to design a wide variety of future electronic devices and to integrate with the high-performance FETs \cite{chuang2016low,schmidt2015electronic,liu2013role,nourbakhsh2016mos2,late2012hysteresis}. Typically, TMDCs-based FETs show an ultrahigh I\textsubscript{on}/I\textsubscript{off} ratio, which is essential for the low-power electronics \cite{radisavljevic2011single,das2014toward}. Moreover, MoS\textsubscript{2} based FETs \cite{yu2017analyzing,perera2013improved,bao2013high,illarionov2018annealing,radisavljevic2011single,zhang2015mos,li2017low,liu2018mos2} exhibit higher carrier mobility, higher I\textsubscript{on}/I\textsubscript{off} ratio, and lower threshold swing, compared with other TMDCs based FETs \cite{liu2013role,iqbal2015high,sik2012transistors}. Therefore, MoS\textsubscript{2} is superior to other materials as a channel layer in the FETs \cite{zhang2012ambipolar} and highly suitable for the optoelectronics \cite{lopez2013ultrasensitive,lembke2015single,pham2019mos2,singh2019flexible,choi2014lateral} and catalysis applications \cite{voiry2013conducting,gali2020electronic,lukowski2013enhanced}. Therefore, MoS\textsubscript{2} based FETs and other advanced devices are leading contenders for the downscaling electronics with a short channel \cite{liu2012channel,ghatak2013observation}, low thickness \cite{lembke2015single,li2014thickness}, small volume \cite{fei2016direct}, fast speed \cite{dhyani2017high,li2016highly}, high sensitivity \cite{kufer2015highly,zhao2017highly}, and lightweight, etc \cite{wang2022evolution,mak2013tightly}.
In addition to a direct bandgap of 1.8 eV \cite{mak2010atomically,kaasbjerg2013acoustic,ganatra2014few,geim2013van,wang2012electronics,butler2013progress}, a film thickness of less than 1 nm \cite{bernardi2013extraordinary,li2015two}, highly sensitive to external pressure \cite{nayak2014pressure}, strain \cite{yue2012mechanical,conley2013bandgap,pena2015single}, and temperature \cite{radisavljevic2013mobility}, the monolayer MoS\textsubscript{2} is advantageous for electronic applications due to its superior electrostatic control of charge density, current even at the transistor scaling limit and insulator/metal transition characteristics under certain conditions \cite{ridolfi2015tight}. 
However, the tensile, compressive, and shear strain allows for extensive tuning of the band gap of monolayer MoS\textsubscript{2} for a variety of applications \cite{li2014strain}.\\
\indent Recent developments \cite{yu2017analyzing,yu2016realization,xiao2014theoretical,ma2014charge} show that electrical and magneto-transport is a rapidly growing research arena. Several theoretical studies have been done to understand the transport properties of MoS\textsubscript{2} \cite{kaasbjerg2013acoustic, ma2014charge, yu2017analyzing, li2016charge, ong2013mobility, yu2016realization, patil2017role}. 
Surprisingly, despite intense research efforts in this field, there is a scarcity of research on the magneto-transport properties of MoS\textsubscript{2}. The magnetic field has a substantial effect on the electrical properties of such materials, with changes quantified by magneto-transport coefficients such as magnetoresistance (MR) and the Hall scattering factor (sometimes referred to as the Hall factor). The MR is an important material property for various applications like data storage, hard drives \cite{DAUGHTON1999334}, senors \cite{56910}, and biomedical applications \cite{RIFE2003209, baselt1998biosensor}. In contrast, the Hall factor knowledge is crucial for measuring carrier concentration and electron mobility. Hence, it is essential to investigate the physical mechanisms governing magneto-transport and develop models capable of producing reliable predictions of the MR and Hall scattering factor. 
\newline \indent In the present work, we advance a carrier transport model based on the Boltzmann transport equation (BTE) \cite{lundstrom2002fundamentals,mandia2021ammcr} and uncover the new important aspects of magneto-transport in monolayer MoS\textsubscript{2}. The current work quantitatively interprets the transport behavior with several substrate materials, identifies the underlying bottlenecks, and provides guidelines for possible subsequent improvements. Furthermore, we show calculations of the MR, the Hall mobility, the Hall scattering factor, electron drift mobility, and the longitudinal conductivity as a function of temperature, carrier density, and magnetic field. The objective of this work is not to seek an extensive overview of all the related identities but rather to explore the various electron transport parameters, understand the underlying device physics, and compare various substrate structures' transport data. Investigations on electrical and magneto-transport properties of the monolayer MoS\textsubscript{2} may, therefore, provide an entirely innovative platform for probing the fundamentals of electron transport in such materials. 

\begin{figure}[!htbp]
	\centering
	\subfigure[\hspace{0.1cm} {}]{\includegraphics[height=0.45\textwidth,width=0.37\textwidth]{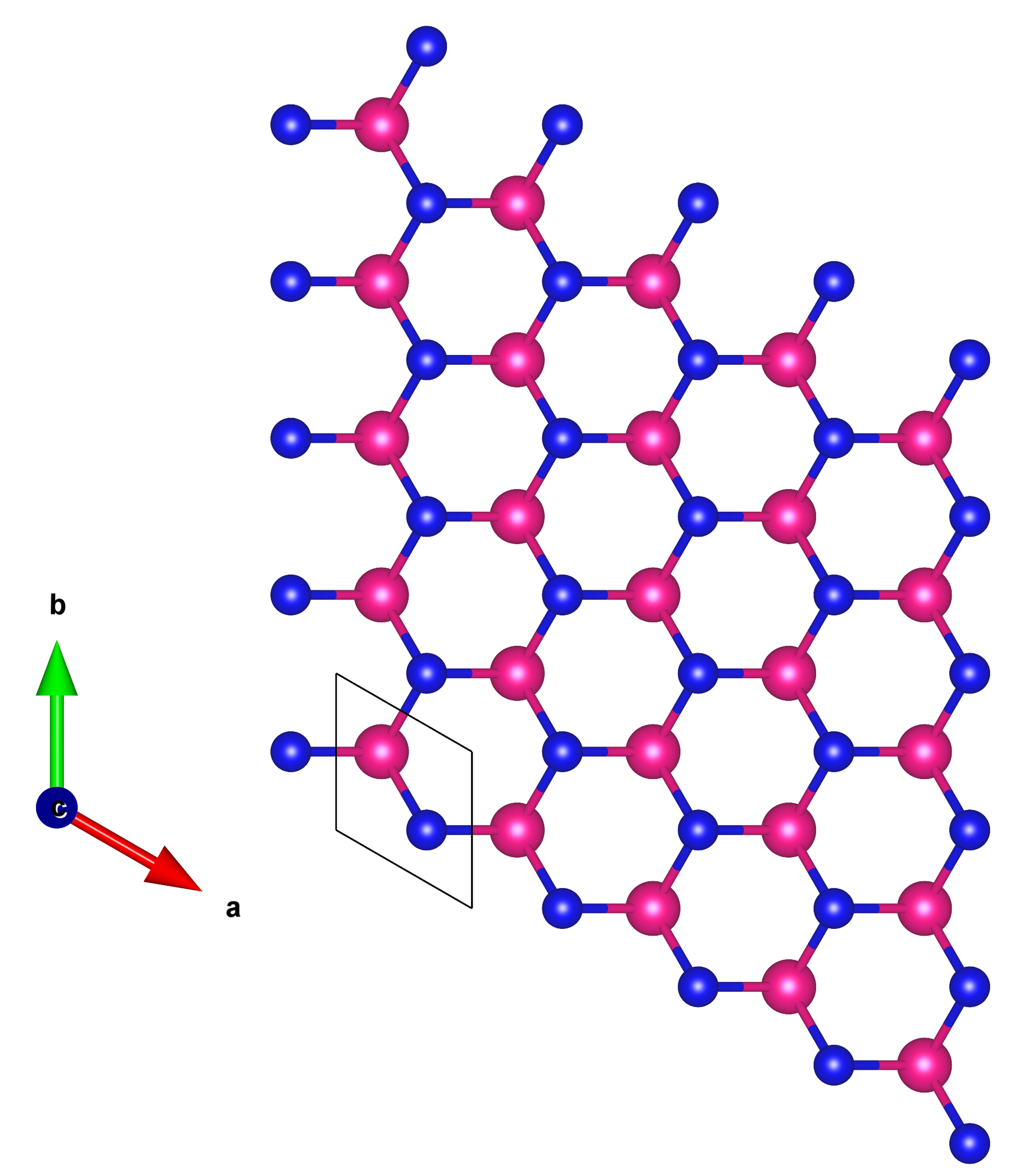}\label{topview}}
	\quad
 \quad
	\subfigure[\hspace{0.1cm} {}]{\includegraphics[height=0.16\textwidth,width=0.40\textwidth]{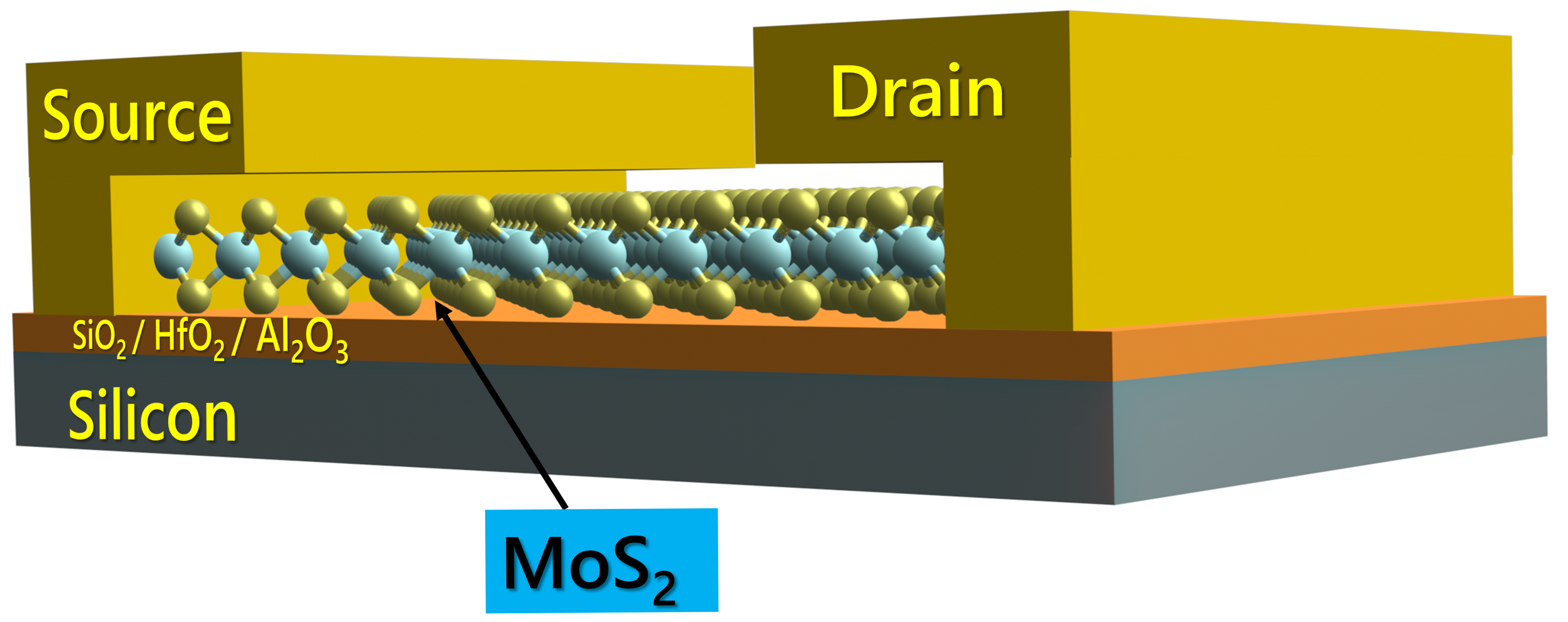}\label{Schematic}}
\caption{ Preliminaries for 2D-MoS\textsubscript{2} (a) typical perspective top view of the  MoS\textsubscript{2} material system (b) schematic illustration of the monolayer MoS\textsubscript{2} based FET devices.}
\label{Preliminaries}
\end{figure}

The manuscript is organized as follows. In Sec. \ref{sec_method}, we review the scattering mechanisms in MoS\textsubscript{2}
and analyze the available monolayer MoS\textsubscript{2} for various substrates, including suspended monolayer MoS\textsubscript{2}. 
In Sec. \ref{results}, we discuss the findings. Finally, in Sec. \ref{conclu}, we draw conclusions and future perspectives from our in-depth analysis. 
Figure \ref{topview} shows the crystal structure of monolayer MoS\textsubscript{2} showing a layer of molybdenum atoms (pink in color) sandwiched between two layers of sulfur atoms (blue). Figure \ref{Schematic} shows a MoS\textsubscript{2} transistor with a self-aligned source/drain terminal and different types of dielectric material.\\

\section{METHODS}
\label{sec_method}
\subsection{Electrical and magneto-transport properties through BTE}
\label{BTE_solution}

\begin{table}[t!]
\caption{\label{table1} Material parameters \cite{yu2017analyzing,kaasbjerg2013acoustic,patil2017role,li2016charge}.}
\begin{ruledtabular}
\begin{tabular}{ccc}
\bf{Quantity} & \bf{Unit}  & \bf{Numerical value} \\ \hline\\
$m^*$ & -& $0.48\: m_0$\\
\\
$E_g$ & eV & $1.8$\\
\\
$\rho_s$ & $Kg/m^{2}$ & $3.1\times 10^{-6}$ \\
\\
$\varepsilon_{_{ion}}^0$ & -& $7.6 \: \varepsilon_0$\\
\\
$\varepsilon_{_{ion}}^\infty$ & -& $7.0 \: \varepsilon_0$\\
\\
$\hbar\:\omega_{Fr}$ & eV & $0.048$\\
\\
\bf{For ADP and PZ}\:: & & \\

$e_{11} $ & C/m & $3\times 10^{-11}$\\
\\

$\Xi_{ADP}^{K,LA \footnote{LA~: Longitudinal\:Acoustic\: Phonons.}}$ &  eV & 2.8\\
\\
$\Xi_{ADP}^{K,TA \footnote{TA~: Transverse\:Acoustic\: Phonons.}}$ &  eV & 1.6\\
\\

$C_{ADP}^{LA}$ &  m/s & $6.7\times10^{3}$\\
\\

$C_{ADP}^{TA}$ &  m/s & $4.2\times10^{3}$\\
\\

\bf{For NPOP}\:: & & \\
\\

$D_1^{K,TA}$ &  eV & 5.9\\
\\
$D_1^{K,LA}$ &  eV & 3.9\\
\\
$D_1^{\Gamma,TO}$ &  eV & 4.0\\
\\
$D_1^{K,TO}$ &  eV & 1.9\\
\\
$D_0^{K,LO\footnote{LO~: Longitudinal\:Optical\: Phonons.}}$ &  eV/m & $2.6\times10^{10}$\\
\\
$D_0^{\Gamma,HP\footnote{HP~: Homopolar\:Optical\: Phonons.}}$ &  eV/m & $4.1\times10^{10}$\\
\\


$\hbar\:\omega_{_{NPOP_1}}^{K,TA}$ &  eV & 0.023\\
\\
$\hbar\:\omega_{_{NPOP_1}}^{K,LA}$ &  eV & 0.029\\
\\
$\hbar\:\omega_{_{NPOP_1}}^{\Gamma,TO}$ &  eV & 0.048\\
\\
$\hbar\:\omega_{_{NPOP_1}}^{K,TO}$ &  eV & 0.047\\
\\

$\hbar\:\omega_{_{NPOP_0}}^{\Gamma,LO}$ &  eV & 0.048\\
\\
$\hbar\:\omega_{_{NPOP_0}}^{\Gamma,HP}$ &  eV & 0.050\\
\\
\end{tabular}
\end{ruledtabular}
\end{table}

The following section provides a brief methodology for solving the BTE. The BTE in terms of electron distribution function $f$ is given by \cite{lundstrom2002fundamentals,singh2007electronic,ferry2016semiconductor}

\begin{equation}
\frac{\partial f(\textbf{k})}{\partial t} + \textbf{v}\cdot\nabla _rf - \frac{e \textbf{E\textsubscript{F}}}{\hbar} \cdot\nabla _kf= \frac{\partial f}{\partial t}\Bigr|_{\substack{coll}},
\label{BTE}
\end{equation}

\noindent where $\textbf{v}$ is the carrier velocity, $e$ is the electronic charge, $\textbf{E\textsubscript{F}}$ is the applied electric field, $\hbar$ is the reduced Planck's constant, and $f$ represents the probability distribution function of electrons in the real and the momentum space as a function of time, $\frac{\partial f}{\partial t}\Bigr|_{\substack{coll}}$ denotes the change in the electron distribution function with time due to the collisions. Under steady-state \bigg($\dfrac{\partial f(\textbf{k})}{\partial t} = 0$\bigg), and spatial homogeneous condition ($\nabla_rf=0$), the BTE (\ref{BTE}) can be rewritten as \cite{mandia2023high} 

\begin{equation}
\begin{split}
\frac{-e {\textbf{E\textsubscript{F}}}}{\hbar}\cdot\nabla _kf = \int \Big[ s(\textbf{k$^\prime$, k})\; f'(1-f)\\
-  s(\textbf{k, k$^\prime$})f(1-f')\Big] d\textbf{k$^\prime$},
\end{split}
\label{BTE1}
\end{equation}

\noindent where $s(\textbf{k}, \textbf{k}^\prime)$ denotes the transition rate of an electron from an initial state $\textbf{k}$ to a final state $\textbf{k}'$.

An iterative method for solving the BTE in the presence of an arbitrary magnetic field was introduced by Rode. The distribution function in such cases can be written as \cite{rode1973theory}  
\begin{equation}
f(\textbf{k})=f_0[E(k)]+xg(k)+yh(k)\:,\\
\label{distribution_func_in_magneticfield}
\end{equation} 

\noindent where, $x$ is the direction cosine measured from $\textbf{E\textsubscript{F}}$ to $\textbf{k}$, $y$ is the direction cosine from $\textbf{B}\times \textbf{E\textsubscript{F}}$ to $\textbf{k}$, B is the applied magnetic field and $k =\bf{|k|}$. The term $g(k)$ is the perturbation to the distribution function due to an applied electric field, and $h(k)$ is the perturbation in the distribution function due to the magnetic field. Substituting Eq. \eqref{distribution_func_in_magneticfield} in the BTE gives a pair of coupled equations that can be solved iteratively \cite{rode1983magnetic,rode1973theory,kumar2023advancing}

\begin{table*}[t!]
\caption{\label{table2} Parameters used for the SO phonon scattering rates \cite{yu2017analyzing,ong2013mobility,zou2010deposition}.}
\begin{ruledtabular}
\begin{tabular}{ccccc}
 & \bf{Unit} & \bf{SiO\textsubscript{2}}  & \bf{Al\textsubscript{2}O\textsubscript{3}} & \bf{HfO\textsubscript{2}} \\ \hline\\

$\varepsilon_{box}^0$& -& 3.90 & 12.53 & 22.00 \\
\\
$\varepsilon_{box}^\infty$& -& 2.50 & 3.20 & 5.03 \\
\\
$\omega_{_{TO1}}$& eV &  0.0556 & 0.04818 & 0.04 \\
$\omega_{_{TO2}}$& eV & 0.1381  & 0.07141 & -  \\
$\omega_{_{LO1}}$&  eV & 0.06257 & 0.05647 & 0.079 \\
$\omega_{_{LO2}}$& eV & 0.15328 & 0.12055 & - \\
\\
\end{tabular}
\end{ruledtabular}
\end{table*}
\begin{equation}
    g_{i+1}(k)=\dfrac{S_i(g_i(k) - \frac {(-e)E\textsubscript{F}}{\hslash}\:(\frac{\partial f_0}{\partial k}) + \beta S_i(h_i(k))}{S_o(k)\:(1+\beta^2)} \:,\\
    \label{perturbation due to E field}
\end{equation} 
\begin{equation}
    h_{i+1}(k)=\dfrac{S_i(h_i(k) + \beta\:\frac {(-e)E\textsubscript{F}}{\hslash}\:(\frac{\partial f_0}{\partial k}) - \beta S_i(g_i(k))}{S_o(k)\:(1+\beta^2)} \:.\\
    \label{perturbation due to magnetic field}
\end{equation} 

The term $\beta$ is a dimensionless quantity and can be expressed as \cite{rode1983magnetic}
\begin{equation}
   \beta=\dfrac {(-e)\nu (k) B}{\hslash k S_o(k)}\:,\\
    \label{beta}
\end{equation} 

\noindent where, the quantities $S_i$ and $S_o$ are the scattering-in and scattering-out operators respectively, and $v =\bf{|v|}$. The above equations show that the perturbation to the distribution function ($g$) due to the applied electric field and ($h$) due to the applied magnetic field are coupled to each other through the factors $\beta$ and the in-scattering $S_i$. It is important to note that such coupled expressions can only be derived using the current iterative method and cannot be obtained using the standard RTA method. 
The electron drift mobility, conductivity, and resistivity can be expressed as \cite{yu2017analyzing}

\begin{equation}
	\mu = - ~ {\dfrac{\sqrt{2m^*}}{\pi~ n~ \hbar^2 E\textsubscript{F}}} ~\bigintssss g(E)~\sqrt{E}~ dE \:,
	\label{mobility}	
\end{equation}

\begin{equation}
	\sigma = n ~ e ~ \mu \:,
	\label{conductivity}	
\end{equation}

\begin{equation}
	\rho = \dfrac {1}{\sigma} \:,
	\label{resistivity}	
\end{equation}

\noindent where $\sigma$ is the conductivity, n is the electron density, and $\mu$ is the electron drift mobility calculated with zero magnetic fields.

In terms of the perturbation, the elements of the conductivity tensor can be expressed as \cite{nag1975galvanomagnetic}   

\begin{equation}
	\sigma_{xx}=- ~ \Bigg( ~ \dfrac{e\sqrt{2m^*}}{\pi \hbar^2 E\textsubscript{F}} \Bigg) ~\bigintssss g(E)~  \sqrt{ E}~ dE \:,
	\label{sigma_xx}	
\end{equation}

\begin{equation}
	\sigma_{xy}=- ~ \Bigg( ~ \dfrac{e\sqrt{2m^*}}{\pi \hbar^2 E\textsubscript{F}} \Bigg) ~\bigintssss h(E)~  \sqrt{ E}~ dE \:.
	\label{sigma_xy}	
\end{equation}

The Hall coefficient $R_H$, the Hall mobility $\mu_H$, and the Hall scattering factor $r_{_H}$  can be expressed as \cite{nag1975galvanomagnetic,rode1983magnetic}.
\begin{equation}
	R_H = \frac{\sigma_{xy}}{{B} ~ [DD]}\:,
	\label{Hall_coeff}	
\end{equation}

where 
\begin{equation}
	DD = {{\sigma_{xx}}^2} + {{\sigma_{xy}}^2} \:,
	\label{DD}	
\end{equation}

\begin{equation}
	\mu_{_H} = \sigma ~ \abs{R_H}\:,
	\label{Hall mobility}	
\end{equation}

\begin{equation}
	r_{_H} = \dfrac{\mu_{_H}}{\mu}\:. 
	\label{Hall factor}	
\end{equation}

The MR can be written as \cite{nag1975galvanomagnetic} 

\begin{equation}
	MR = \frac{\rho_{xx} - \rho_{xx}(0)}{\rho_{xx}(0)}= \Bigg[\dfrac{\sigma_{xx} ~ \sigma(0)}{DD}\Bigg] -1 \:,\label{MR}	
\end{equation}
\noindent where $\rho(0)$ and $\sigma(0)$ represent the resistivity and conductivity at zero magnetic fields, respectively, and $\rho_{xx}$ represents the longitudinal resistivity and is given by \cite{ferry2016semiconductor}
\begin{equation}
	\rho_{xx} = \dfrac{\abs{\sigma_{xx}}}{DD}\:.
 \label{Longitudinal_resistivity}	
\end{equation}

\subsection{Computational details}
\subsubsection*{\bf\emph{1. \hspace{0.2cm}{Remote charge impurity scattering}}}

The scattering rate for remote charge impurity (CI) scattering can be written as \cite{patil2017role,yu2017analyzing} 

\begin{equation}
	\frac{1}{\tau_{_{CI}}}=\frac{N_{_I}~e^4 ~ m^* }{8 \pi \hbar^3 K_{avg}^2} \int_0 ^{2 \pi} \frac{1}{2 k^2} \frac{1}{\varepsilon_{2D}^2(q,T)} ~ d\theta,
	\label{charge_impurity}	
\end{equation}
where $m^*$ is the electron effective mass, $N_{_I}$ is the remote impurity concentration which is taken as $1 \times 10^{12}~cm^{-2}$ in our simulations, and $K_{avg}$ is the effective dielectric constant of the surroundings. 
For supported monolayers (MLs), $K_{avg} = ((1+\varepsilon_{box}^{0})/2)$, where $\varepsilon_{box}^{0}$ represents the substrate dielectric constant. The term $\varepsilon_{2D}(q,T)$ is the general static dielectric function and is given by \cite{ong2013mobility}

\begin{equation}
	\varepsilon_{2D}(q,T) = 1 + e^2 G_q \Pi(q,T,E_F),
	\label{static dielectric fun}	
\end{equation}
where $G_q = [(1 + \varepsilon_0 \varepsilon_{box}^0)q]^{-1}$. Here $\varepsilon_0$ is the vacuum permittivity, and $q$ represents the change in wave vector due to scattering. For elastic scattering mechanisms $q = 2k sin(\theta/2)$, here $\theta$ is the angle between the incoming and outgoing wave vector. The quantity $\Pi(q,T,E_F)$ is the static polarizability that represents the polarization charge screening of the coulomb impurities and depends on the temperature and carrier density. At finite $q$, this charge polarizability can be written as

\begin{equation}
	\Pi~(q,T,E_F)=\int_0^\infty  \frac{\Pi~(q,0,\mu)}{4~k_B T ~cosh^2 \Big(\frac{E_F-\mu}{2k_B T}\Big)}~ d\mu,
	\label{polarizability}	
\end{equation}
where $\Pi~(q,0,\mu)$ = $[-\textsl{g} m^{*}/(2 \pi \hbar^2)]$\{1 - $\Theta (q - 2 k_F ) [1-(2 k_F /q)^2]^\frac{1}{2}$\} with $k_F = \sqrt{2 m^{*}~\mu} /\hbar$. Here, $\textsl{g}$ is the valley-spin degeneracy ($\textsl{g}=4$). 
\vspace{-0.5em}
\subsubsection*{\bf\emph{2. \hspace{0.2cm}{Acoustic deformation potential scattering}}}
The scattering rate caused by the deformation potential of the acoustic phonons (ADP) can be expressed as \cite{li2016charge,yu2017analyzing}

\begin{equation}
\frac{1}{\tau_{ADP}} = \frac{\Xi_{ADP}^2~k_B T~m^* }{2 \pi \hbar^3 \rho_s ~ c_{ADP}^2} \int _0 ^{2 \pi} \frac{(1-cos ~ \theta)}{\varepsilon_{2D}^2} ~ d\theta,
 \label{ADP}	
\end{equation}

\noindent where, $T$ and $k_B$ are the temperatures and Boltzmann constant, respectively. $\Xi_{ADP}$ is the acoustic phonon (LA or TA) deformation potential, $\rho_s$ is the sheet mass density of a single layer of MoS\textsubscript{2}, $\theta$ is the elastic scattering angle from the initial momentum $k$ to the final momentum $k^\prime$, and $c_{ADP}$ is the acoustic phonon speed.
\vspace{-0.5em}
\subsubsection*{\bf\emph{3. \hspace{0.2cm}{Piezoelectric scattering}}}
Under long-wavelength approximation, the high-temperature relaxation rate for the piezoelectric (PZ) scattering can be written as \cite{li2016charge,kaasbjerg2013acoustic} 

\begin{equation}
\frac{1}{\tau_{_{PZ}}} = \frac{1}{\tau_{_{ADP}}} \times \frac{(e_{_{11}}e/\varepsilon_{_0})^2}{2~\Xi_{_{ADP}}^2} ,
 \label{PZ}	
\end{equation}
where $e_{_{11}}$ is the piezoelectric constant.  

\subsubsection*{\bf\emph{4. \hspace{0.2cm}{Non-polar optical phonon scattering}}}
The scattering rate of the zeroth and first order of the non-polar optical phonon (NPOP) is given by \cite{li2016charge}  
\begin{equation}
\begin{split}
&S^{i\pm}_{NPOP_0}(E) =\dfrac{1}{4\pi \hbar^2 \rho_{_S} \omega_{_{NPOP_0}}}\Bigg[D_{0}^2 ~m^* \Bigg(\dfrac{1}{2}\pm \dfrac{1}{2}\\
& +N_{NPOP_0} \Bigg)\Bigg] \bigintss_0^{2 \pi}\frac{1}{\varepsilon_{2D}^2}\Bigg(\dfrac{-k^{\prime}}{k}\Bigg)cos\theta ~d\theta \:,
\end{split}
 \label{NPOP_0i}	
\end{equation}

\begin{equation}
\begin{split}
&S^{o\pm}_{NPOP_0}(E) =\dfrac{1}{4\pi \hbar^2 \rho_{_S} \omega_{_{NPOP_0}}}\Bigg[D_{0}^2 ~m^* \Bigg(\dfrac{1}{2}\pm \dfrac{1}{2}\\
&+N_{NPOP_0} \Bigg)\Bigg]\bigintss_0^{2 \pi}\frac{1}{\varepsilon_{2D}^2} ~d\theta \:,
\end{split}
 \label{NPOP_0o}	
\end{equation}

\begin{equation}
\begin{split}
&S^{i\pm}_{NPOP_1}(E) =\dfrac{1}{4\pi \hbar^2 \rho_{_S} \omega_{_{NPOP_1}}}\Bigg[D_{1}^2 ~m^* \Bigg(\dfrac{1}{2}\pm \dfrac{1}{2}\\
&+N_{NPOP_1} \Bigg)\Bigg]\bigintss_0^{2 \pi}\dfrac{1}{\varepsilon_{2D}^2}~q^2\Bigg(\dfrac{-k^{\prime}}{k}\Bigg)cos\theta ~d\theta \:,
\end{split}
 \label{NPOP_1sti}	
\end{equation}

\begin{equation}
\begin{split}
&S^{o\pm}_{NPOP_1}(E) =\dfrac{1}{4\pi \hbar^2 \rho_{_S} \omega_{_{NPOP_1}}}\Bigg[D_{1}^2 ~m^* \Bigg(\dfrac{1}{2}\pm \dfrac{1}{2}\\
&+N_{NPOP_1} \Bigg)\Bigg]\bigintss_0^{2 \pi}\frac{1}{\varepsilon_{2D}^2}~ q^2 ~d\theta \:,
\end{split}
 \label{NPOP_1sto}	
\end{equation}
where $ q=2k sin (\theta /2)$ is the scattering vector, $D_{0}$ and $D_{1}$ denotes the zeroth-and first-order optical deformation potential, respectively, $N_{NPOP}$ is the number of phonons follows the Bose-Einstein distribution because phonon scattering depends highly on temperature, and can be expressed as $N_{NPOP}=1/[\exp(\hbar\omega_{NPOP}/k_BT)-1]$ and $\omega_{NPOP_0}$ and $\omega_{NPOP_1}$ are the frequency of the NPOP mode for zeroth and first order, respectively.

\vspace{-1em}
\subsubsection*{\bf\emph{5. \hspace{0.2cm}{Polar optical phonon scattering}}}
Under long wavelength limits, the intravalley polar LO (Froehlich) phonons are allowed to couple with the electron gas and undergo screening. In such cases, the Froehlich optical phonon in-scattering and out-scattering rate for emission (+) or absorption (-) of polar optical phonons (POP) can be written as \cite{yu2017analyzing}

\begin{equation}
\begin{split}
&S^{i\pm}_{Fr}(E) =\frac{e^2\omega_{Fr}m^*}{8\pi\hbar^2}\Bigg(\frac{1}{2}\pm \frac{1}{2}+N_{Fr} \Bigg)\bigintss_{0}^{2\pi} \dfrac{1}{q}\Bigg(-\dfrac{k^{\prime}}{k}\Bigg)\\
&cos\theta ~d\theta \Bigg( \frac{1}{\varepsilon_{ion}^\infty + \varepsilon_{el}(q)} -\frac{1}{\varepsilon_{ion}^0 + \varepsilon_{el}(q)} \Bigg)\erfc\Bigg(\dfrac{q\sigma_s}{2}\Bigg)^2\:,
\end{split}
 \label{POP_in}	
\end{equation}

\begin{equation}
\begin{split}
S^{o\pm}_{Fr}(E)=\frac{e^2\omega_{Fr}m^*}{8\pi\hbar^2}\Bigg(\frac{1}{2}\pm \frac{1}{2}+N_{Fr} \Bigg)\int_{0}^{2\pi}\dfrac{1}{q}~ d\theta \\ 
\Bigg( \frac{1}{\varepsilon_{ion}^\infty + \varepsilon_{el}(q)} -\frac{1}{\varepsilon_{ion}^0 + \varepsilon_{el}(q)} \Bigg)\erfc\Bigg(\dfrac{q\sigma_s}{2}\Bigg)^2\:,
\end{split}
 \label{POP_out}	
\end{equation}
where $k$ is initial wave vector, $k^\prime$ is the final wave vector given by $k^\prime=\sqrt{k^2\mp2m^*\omega_{_{Fr}}/\hbar}$, and change in vector $q=\sqrt{k^2+{k^{\prime}}^2-2kk^\prime cos\theta}$. $\omega_{Fr}$ is the frequency of the polar optical phonon. $\varepsilon_{ion}^\infty$ 
 and $\varepsilon_{ion}^0$ are the ionic part of the optical and the static permittivity of MoS\textsubscript{2}, $\sigma_s$ is the sheet thickness which is taken as $4.41 \times 10^{-8}~ cm$, and erfc is the complementary error function. 
 The electronic part of the dielectric function can be written as $\varepsilon_{el}(q)=-\dfrac{e^2}{2q}\Pi(q,T,E_F)$.
\vspace{-0.5em}
\subsubsection*{\bf\emph{6. \hspace{0.2cm}{Remote interface or surface optical phonon scattering}}}

To determine the remote interaction between the electrons and the substrate surface optical (SO) phonons, the coupling coefficient can be written as \cite{yu2017analyzing} 
\begin{equation}
M_q =\Bigg[ \frac{e^2 \hbar \omega_{_{SO}}}{\Omega q}      \Bigg( \frac{1}{\varepsilon_{tot,SO}^{\infty} + \varepsilon_{el} (q)}  -\frac{1}{\varepsilon_{tot,SO}^{0} + \varepsilon_{el} (q)} \Bigg)\Bigg ]^{\dfrac{1}{2}}\:,
 \label{remote interaction}	
\end{equation}
where, $\omega_{_{SO}}$ is the SO phonon energy, and $\varepsilon_{_{tot,SO}}^{\infty}$ and $\varepsilon_{_{tot,SO}}^{0}$ are the optical and static dielectric response of the interface, respectively which can be described as 

\begin {equation}
\varepsilon_{_{tot, SO}}^{\infty}= \dfrac{1}{2} \Bigg[\varepsilon_{box}^{\infty} \Big(\frac{\omega_{LO_2}^2 - \omega_{SO}^2}{\omega_{TO_2}^2 - \omega_{SO}^2} \Big)+ \varepsilon_0 \Bigg] \:,
\end{equation}
\begin{equation}
\varepsilon_{_{tot, SO}}^{0}= \dfrac{1}{2} \Bigg[\varepsilon_{box}^{\infty}\Bigg(\dfrac{\omega^2_{LO_1}}{\omega^2_{TO_1}}\Bigg) \Big(\frac{\omega_{LO_2}^2 - \omega_{SO}^2}{\omega_{TO_2}^2 - \omega_{SO}^2} \Big) + \varepsilon_0\Bigg].
\end{equation}

For HfO\textsubscript{2}, which has only one LO and TO mode, the expression can be written as 

\begin {equation}
\varepsilon_{_{tot, SO}}^{\infty}= \dfrac{1}{2} \Bigg[\varepsilon_{box}^{\infty} + \varepsilon_0 \Bigg]\:, 
\end{equation}
\begin{equation}
\varepsilon_{_{tot, SO}}^{0}= \dfrac{1}{2} \Bigg[\varepsilon_{box}^{\infty}\Bigg(\dfrac{\omega^2_{LO}}{\omega^2_{TO}}\Bigg) + \varepsilon_0\Bigg].
\end{equation}

The frequency response of the dielectric is described by the dielectric function of the substrate, denoted by $\varepsilon_{box}(\omega)$ \cite{yu2017analyzing} 

\begin{equation}
\begin{split}
\varepsilon_{box}(\omega)=\varepsilon_{box}^\infty + (\varepsilon_{box}^i - \varepsilon_{box}^\infty) \Bigg[\frac{\omega^2_{TO2}}{\omega^2_{TO2}-\omega^2} \Bigg]\\
+ ~(\varepsilon_{box}^0 - \varepsilon_{box}^i) \Bigg[\frac{\omega^2_{TO1}}{\omega^2_{TO1}-\omega^2}\Bigg] ,
\end{split}
 \label{dielectric fun}	
\end{equation}
where  $\varepsilon_{box}^i ~, \varepsilon_{box}^0$ and $\varepsilon_{box}^\infty$ are the intermediate, static, and optical dielectric response of the substrate, respectively. The term $\omega_{_{TO1}}$ and $\omega_{_{TO2}}$ represents the angular frequencies of the TO phonons, where $\omega_{_{TO1}} ~ < ~\omega_{_{TO2}}$. Rewriting $\varepsilon_{box}(\omega)$ in the generalized Lyddane-Sachs-Teller form as \cite{yu2017analyzing} 

\begin{equation}
\varepsilon_{box}(\omega)=\varepsilon_{box}^\infty ~ \Bigg[\frac{\omega^2_{LO1}-\omega^2}{\omega^2_{TO1}-\omega^2}\Bigg] \\
\Bigg[\frac{\omega^2_{LO2}-\omega^2}{\omega^2_{TO2}-\omega^2}\Bigg].
 \label{Lyddane-Sachs-Teller}	
\end{equation}

The dielectric function of HfO\textsubscript{2}, which has only one TO mode can be written as \cite{yu2017analyzing}  

\begin{equation}
\begin{split}
\varepsilon_{box}(\omega)=\varepsilon_{box}^\infty + (\varepsilon_{box}^0 - \varepsilon_{box}^\infty) \Bigg[\frac{\omega^2_{TO}}{\omega^2_{TO}-\omega^2} \Bigg]\\
= \varepsilon_{box}^{\infty} \Bigg[\frac{\omega^2_{LO}-\omega^2}{\omega^2_{TO}-\omega^2}\Bigg].
\end{split}
 \label{dielectric fun for HfO2}	
\end{equation}

The SO phonon frequencies $\omega_{SO_1}$ and $\omega_{SO_2}$ can be determined from the roots of the equation $\varepsilon_{box}(\omega) + \varepsilon_0 = 0 $.  

The in-scattering and out-scattering rate due to SO phonons can be written as \cite{yu2017analyzing,li2016charge}

\begin{equation}
\begin{split}
S^{i\pm}_{SO}(E) &=\frac{e^2\omega_{_{SO}}m^*}{8\pi\hbar^2}\Bigg(\frac{1}{2}\pm \frac{1}{2}+N_{SO} \Bigg)\bigintss_{0}^{2\pi} \dfrac{1}{q} \Bigg(-\dfrac{k^{\prime}}{k}\Bigg)\\
&cos\theta ~d\theta \Bigg( \frac{1}{\varepsilon_{tot,SO}^\infty + \varepsilon_{el}(q)} -\frac{1}{\varepsilon_{tot,SO}^0 + \varepsilon_{el}(q)} \Bigg)\:,
\end{split}
 \label{SOP_in}	
\end{equation}

\begin{equation}
\begin{split}
S^{o\pm}_{SO}(E)=\frac{e^2\omega_{_{SO}}m^*}{8\pi\hbar^2}\Bigg(\frac{1}{2}\pm \frac{1}{2}+N_{SO} \Bigg)\int_{0}^{2\pi}\dfrac{1}{q}~ d\theta \\ 
\Bigg( \frac{1}{\varepsilon_{tot,SO}^\infty + \varepsilon_{el}(q)} -\frac{1}{\varepsilon_{tot,SO}^0 + \varepsilon_{el}(q)} \Bigg)\:.
\end{split}
 \label{SOP_out}	
\end{equation}

The material parameters used for various scattering rate calculations are given in Table \ref{table1} and \ref{table2}.

\section{RESULTS AND DISCUSSION}
\label{results}

\begin{figure}[!htbp]
	\centering
	\subfigure[\hspace{0.1cm} {B = 0 T and n = $1\times10^{11} ~ cm^{-2}$}]{\includegraphics[height=0.35\textwidth,width=0.45\textwidth]{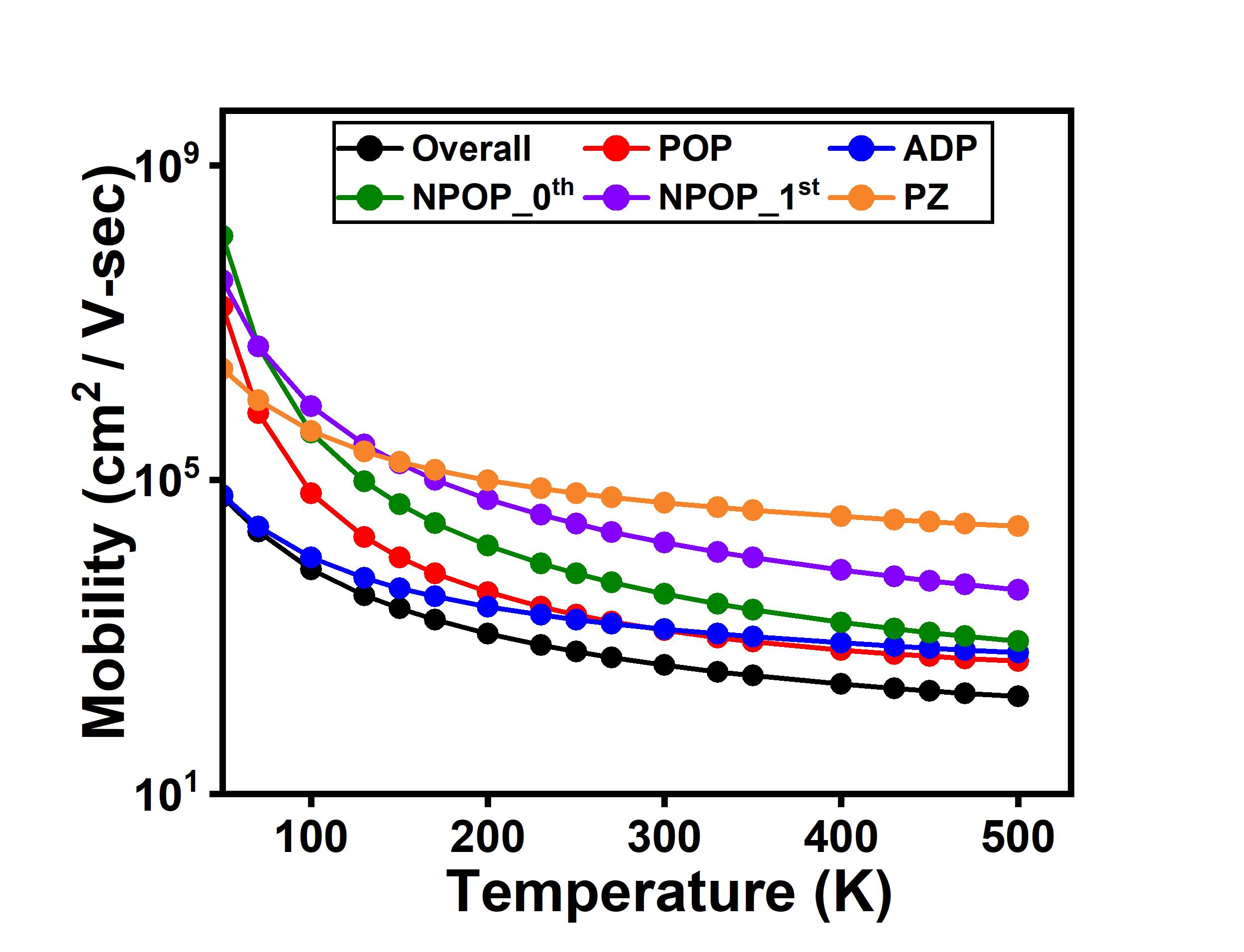}\label{B_0T_1e15}}
	\quad
	\subfigure[\hspace{0.1cm} {B = 0 T and n = $1\times10^{14} ~ cm^{-2}$}]{\includegraphics[height=0.35\textwidth,width=0.45\textwidth]{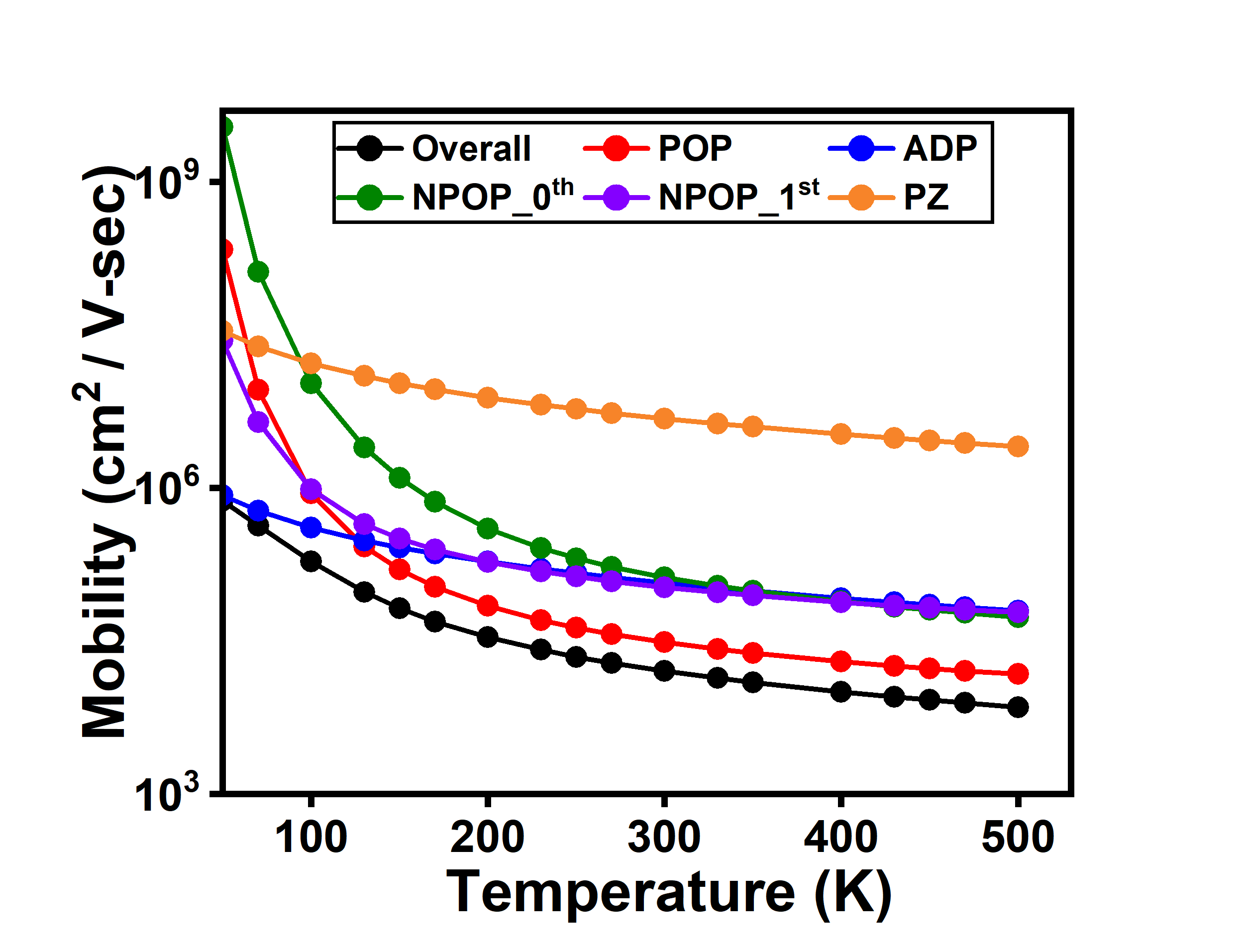}\label{B_0T_1e18}}
\caption{ Contribution to the electron drift mobility by various scattering mechanisms in monolayer MoS\textsubscript{2} (a) temperature dependence of the electron drift mobility for carrier density of $1\times10^{11} ~ cm^{-2}$ (b) temperature dependence of the electron drift mobility for carrier density of $1\times10^{14} ~ cm^{-2}$.}
\label{contribution_to_the_mobility_without_substrate}
\end{figure}

We compute the electron drift mobility, Hall mobility, conductivity, Hall scattering factor, MR, and element of the conductivity tensor in monolayer MoS\textsubscript{2} as a function of temperature, carrier density, and magnetic field. 

\begin{figure}[!t]
	\centering	{\includegraphics[height=0.35\textwidth,width=0.45\textwidth]{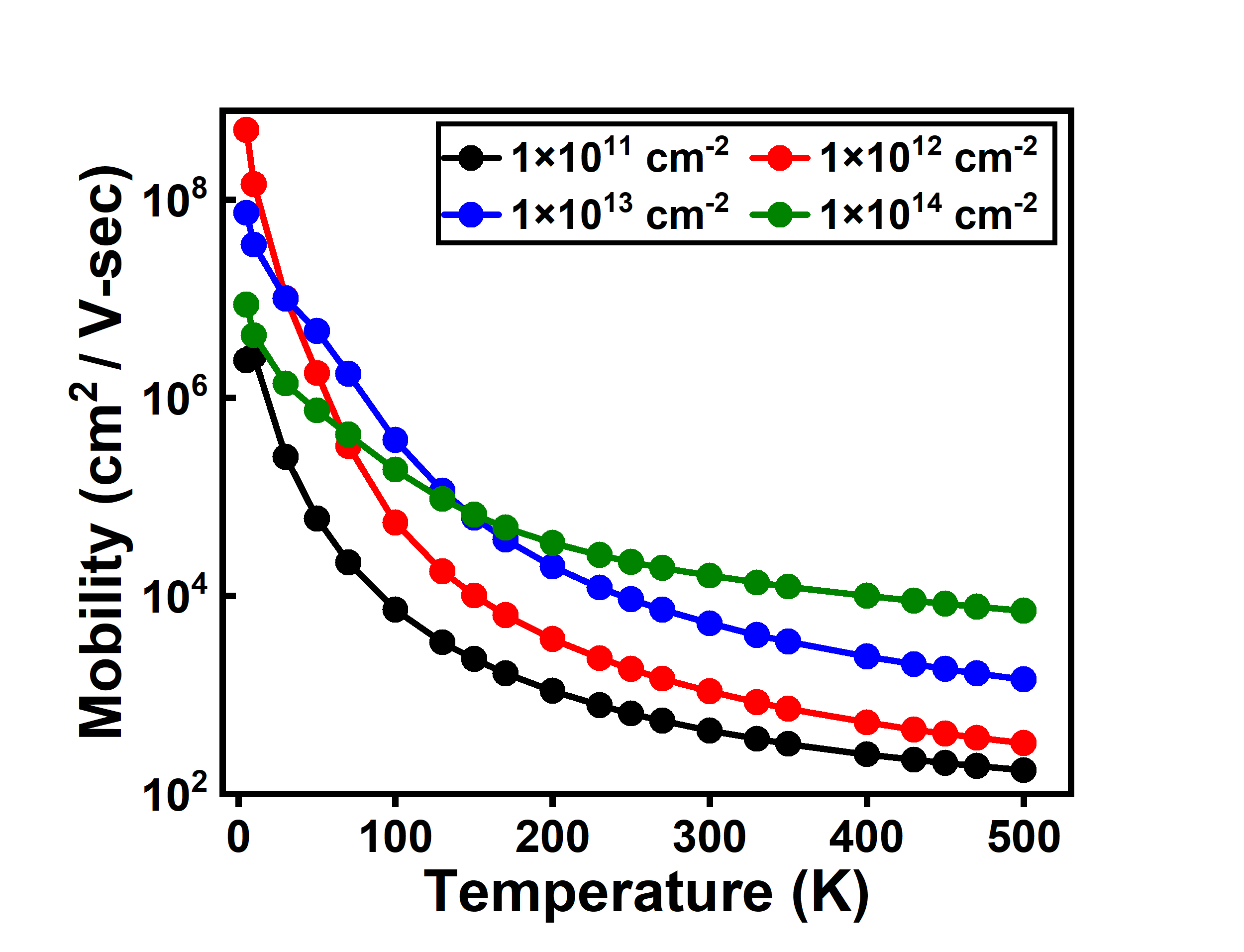}}
	\quad
	\caption{Doping dependence of the electron drift mobility as a function of temperature at B = 0~T in the suspended monolayer MoS\textsubscript{2}. }
	\label{B_0T_without_substrate_drift_mobility}
\end{figure}

\indent In Fig.\ref{contribution_to_the_mobility_without_substrate}, we show the contribution to the electron drift mobility by various scattering mechanisms. Herein, we demonstrate the mobility contribution at two different carrier densities. The ADP and the POP scattering are the most dominant scattering mechanisms in suspended monolayer MoS\textsubscript{2}. At low temperatures, acoustic phonon-dominated transport causes a significant shift in the temperature dependence of electron drift mobility. The POP scattering mechanism becomes dominant at higher temperatures and carrier densities. At lower carrier density, the contribution from the zeroth order NPOP (i.e., $NPOP\_0^{th}$) scattering mechanism is more than the first order NPOP (i.e., $NPOP\_1^{st}$) scattering mechanism. However, contribution order changes significantly at higher densities, as shown in Fig. \ref{contribution_to_the_mobility_without_substrate}. 
\begin{figure}[!t]
	\centering	{\includegraphics[height=0.35\textwidth,width=0.45\textwidth]{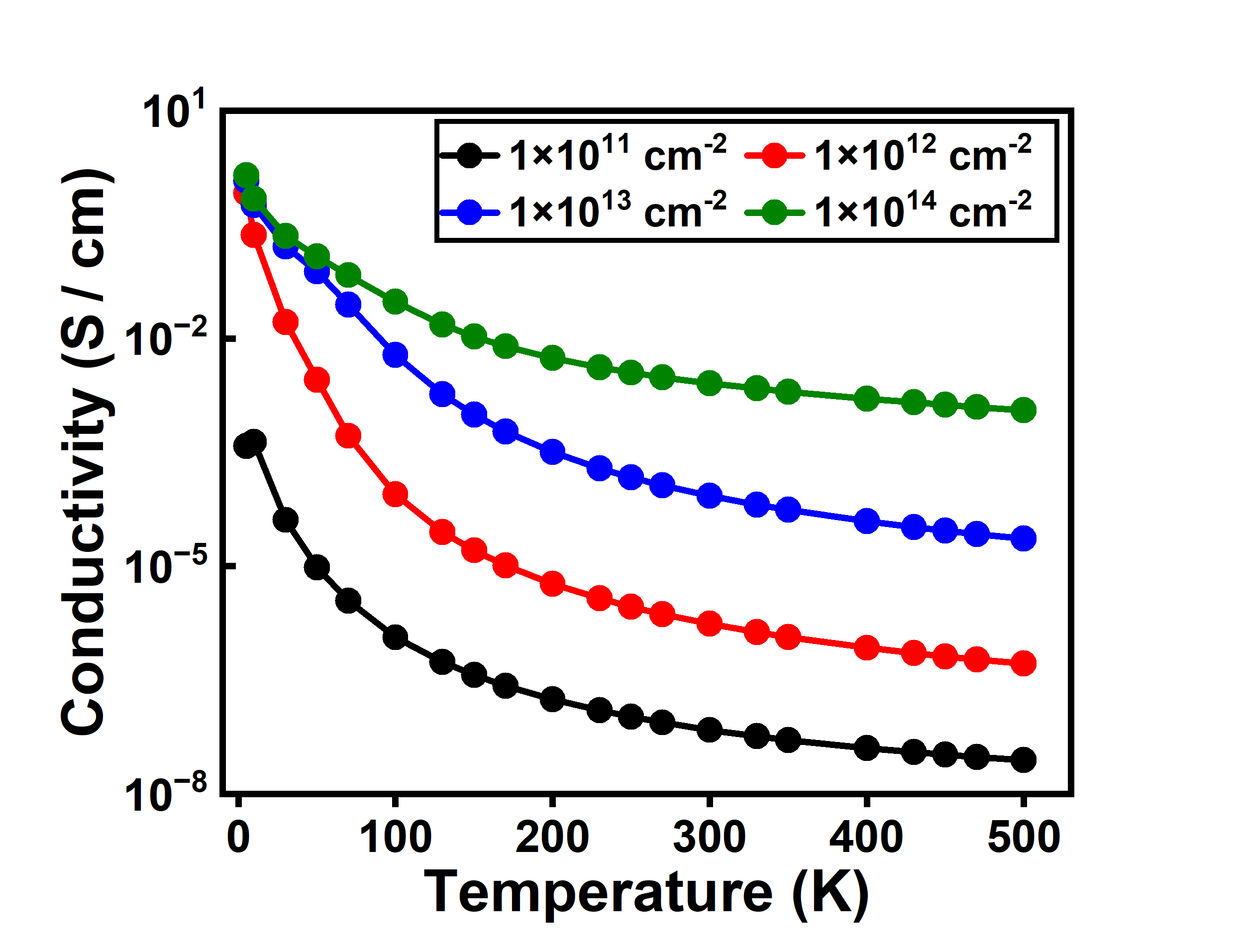}}
	\quad
	\caption{Conductivity in the suspended monolayer MoS\textsubscript{2} as a function of temperature and doping density at B=0 T.}
	\label{Conductivity}
\end{figure}
\begin{figure}[!t]
	\centering	{\includegraphics[height=0.35\textwidth,width=0.45\textwidth]{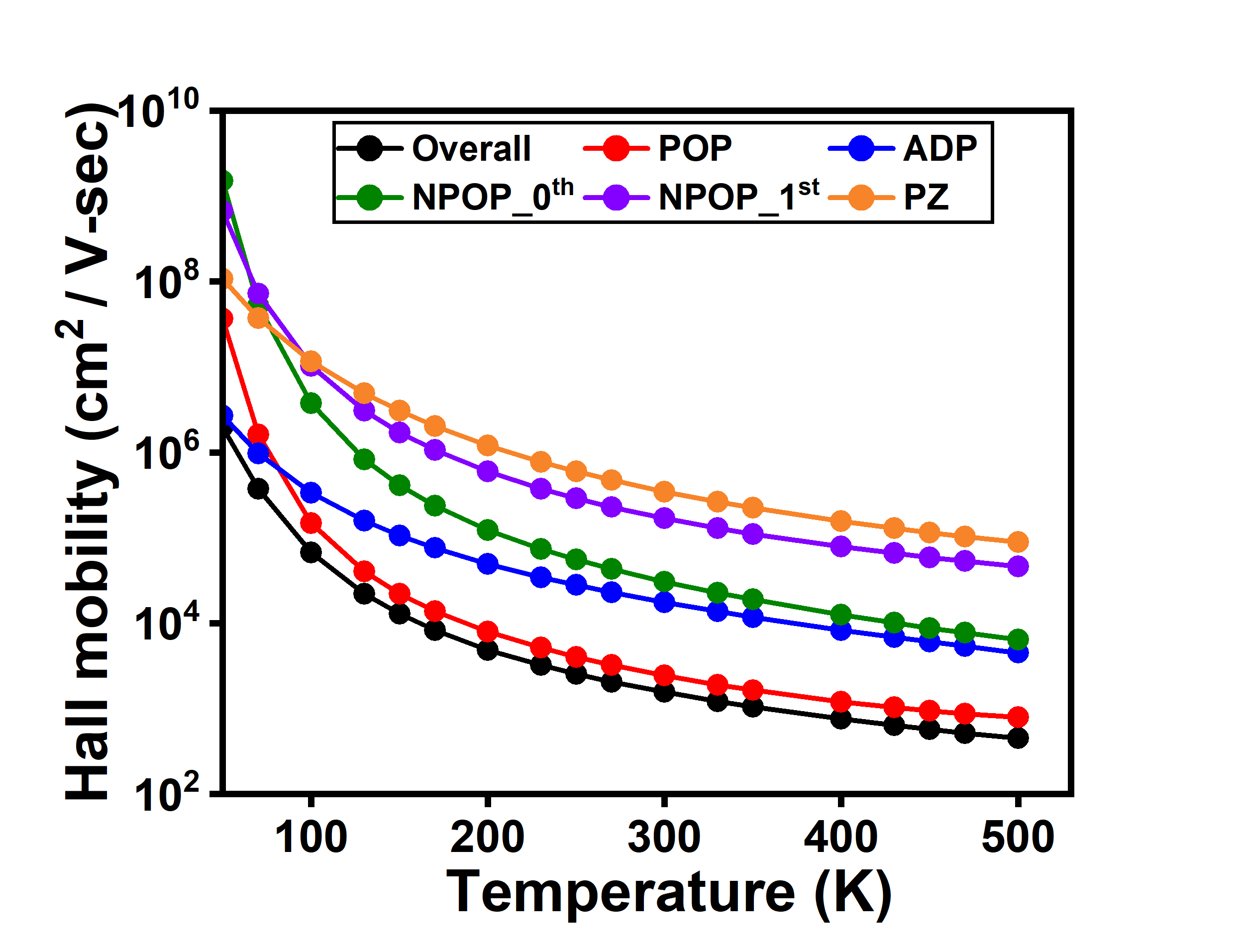}}
	\quad
	\caption{Contribution to the Hall mobility by various scattering mechanisms in the suspended monolayer MoS\textsubscript{2} for a carrier density of $1\times 10^{12}~cm^{-2}$ at B = 0.3 T.}
	\label{Contribution to the Hall mobility}
\end{figure}

In Fig. \ref{B_0T_without_substrate_drift_mobility}, we show the temperature and doping dependence of the electron drift mobility in a suspended monolayer MoS\textsubscript{2} at zero magnetic fields. The increase in temperature increases the phonon scattering rate, consequently reducing the mobility of the carriers. Above $ 200$ K, the electron drift mobility monotonically increases with carrier concentration. Therefore, it is evident that a higher density of carriers is required to achieve higher carrier mobility. 
\begin{figure}[!t]
	\centering
	\subfigure[\hspace{0.1cm} {}]{\includegraphics[height=0.35\textwidth,width=0.45\textwidth]{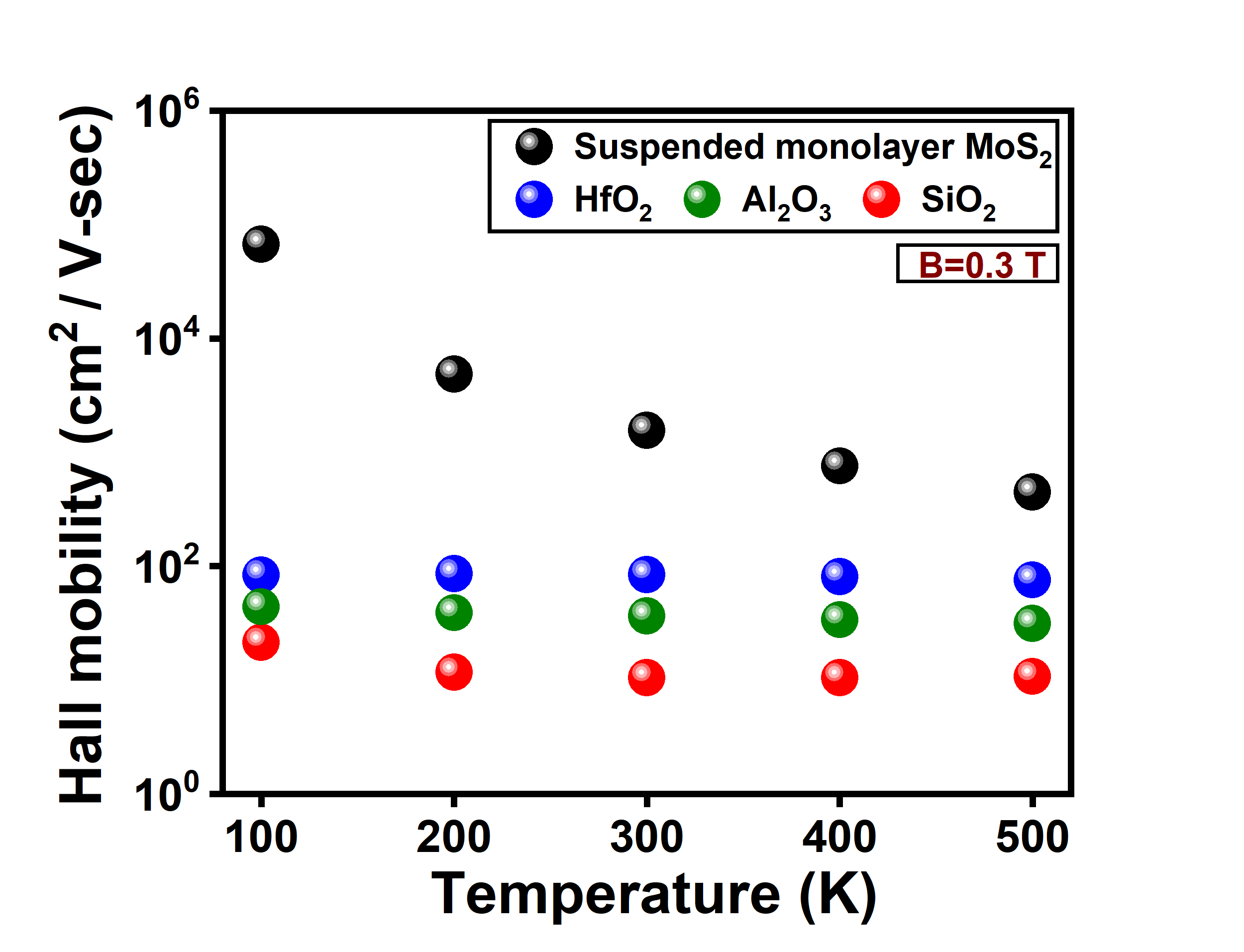}\label{Hall mobility vs temp at B 0.3 T and 1e16 doping}}
		\quad
	\subfigure[\hspace{0.1cm} {}]{\includegraphics[height=0.35\textwidth,width=0.45\textwidth]{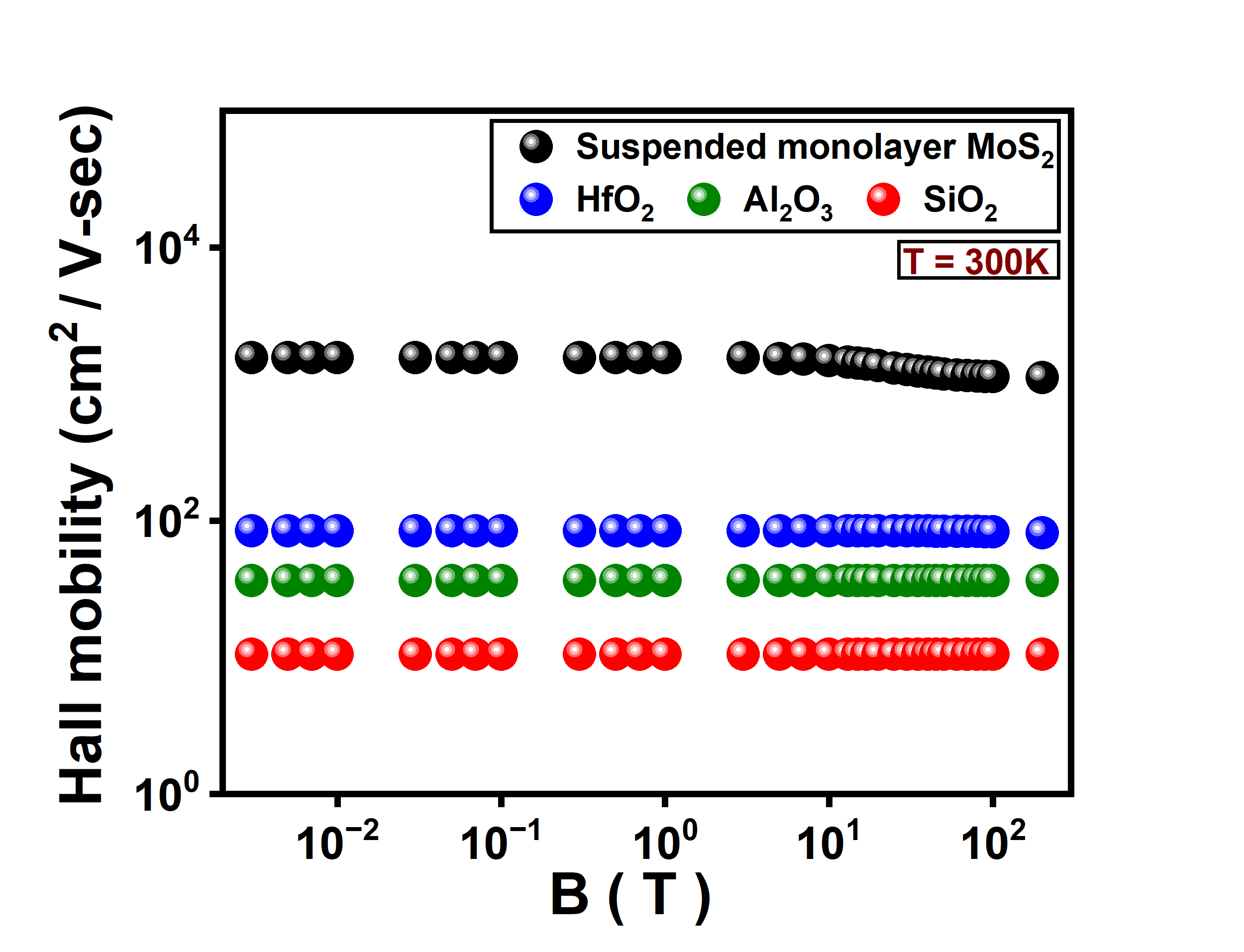}\label{Hall mobility vs magnetic field at T 300 K and 1e16 doping}}

\caption{Variation of the Hall mobility in the monolayer MoS\textsubscript{2} (a) temperature dependence of the Hall mobility  for a carrier density of $1\times 10^{12}~cm^{-2}$ at B = 0.3 T\hspace{0.1cm} (b) the Hall mobility as a function of the magnetic field for a carrier density of $1\times 10^{12}~cm^{-2}$ at T= 300 K.}
\label{Hall mobility with and without substrate}
\end{figure}

In Fig. \ref{Conductivity}, we show the conductivity in a suspended monolayer MoS\textsubscript{2} as a function of temperature and doping density, and the magnetic field is kept at zero. The conductivity increases as the doping density increases since conductivity is directly proportional to the carrier density and decreases as we raise the temperatures due to an increase in phonon scattering. Due to its low electrical conductivity, its power factor in thermoelectric applications is severely constrained.

The contribution to the Hall mobility by various scattering mechanisms as a function of temperature is shown in Fig. \ref{Contribution to the Hall mobility}. Above 100 K, the overall mobility is governed by the POP scattering mechanism, which becomes dominant. 
Compared to other scattering mechanisms, the PZ scattering mechanism contributes the least. At very low temperatures, the contributions from the zeroth and first orders of the NPOP scattering mechanism are nearly equal, but as temperature increases, there is a noticeable change in the contribution to the Hall mobility, highlighting the importance of the contribution to the mobility plot.

\begin{figure}[!t]
	\centering	{\includegraphics[height=0.35\textwidth,width=0.45\textwidth]{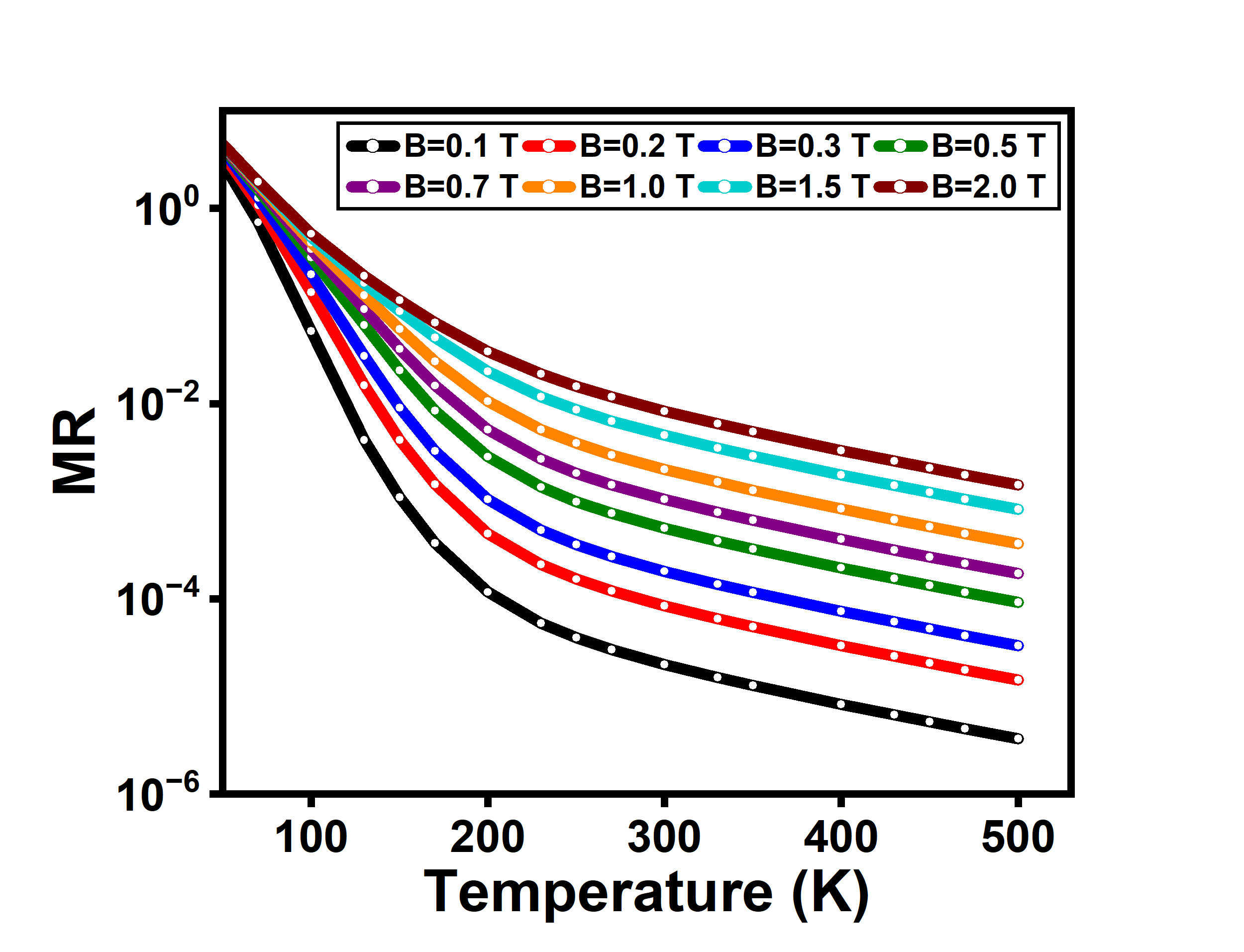}}
	\quad
	\caption{The magnetic field and temperature dependence of the MR in the suspended monolayer MoS\textsubscript{2} for a carrier density of $1\times 10^{12}~cm^{-2}$.}
	\label{MR without substrate}
\end{figure}

\begin{figure}[!t]
	\centering	{\includegraphics[height=0.35\textwidth,width=0.45\textwidth]{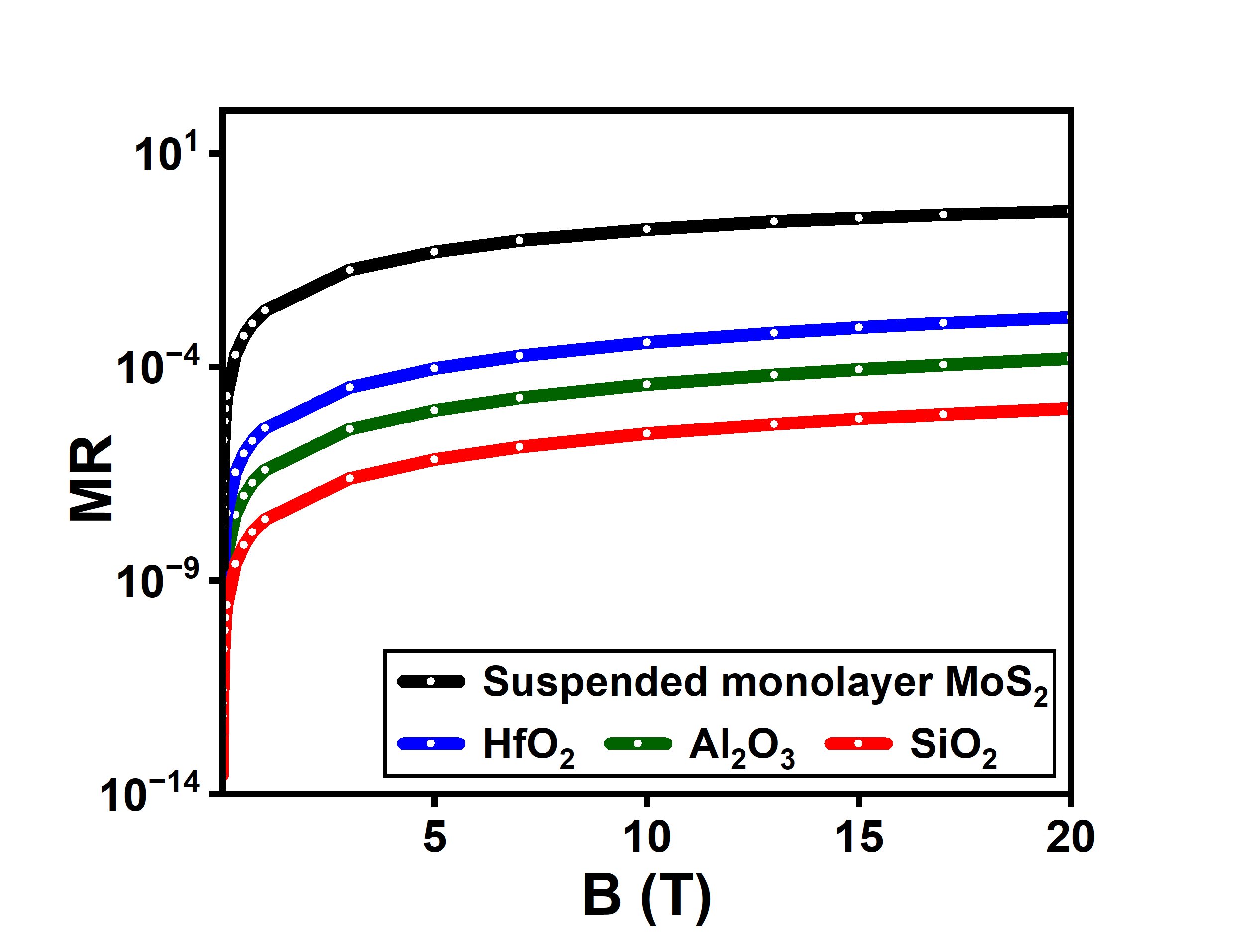}}
	\quad
	\caption{The magnetic field dependence of the MR with various dielectric environments in the monolayer MoS\textsubscript{2} at room temperature for a carrier density of $1\times 10^{12}~cm^{-2}$.}
	\label{MR in substrate less MoS2}
\end{figure}

In Fig. \ref{Hall mobility with and without substrate}, we plot the Hall mobility by combining the effect of various scattering mechanisms for three different dielectric environments with temperature and magnetic fields. We found that the Hall mobility of MoS\textsubscript{2} decreases when it is fabricated over a substrate due to scattering from remote impurity and remote polar optical phonons of the substrate. Of all the substrates, Hall mobility is highest with HfO\textsubscript{2} substrate. Hence, HfO\textsubscript{2} is the best choice to achieve higher mobility since it is difficult to fabricate a suspended MoS\textsubscript{2} layer. The Hall mobility as a function of temperature is shown in Fig. \ref{Hall mobility vs temp at B 0.3 T and 1e16 doping}. For all three dielectric materials, mobility remains almost constant with temperature variation. Remote impurity scattering has a higher scattering rate at lower temperatures, while phonon scattering has a higher scattering rate at higher temperatures. With the increase in temperature, lattice phonon scattering rates increase, but simultaneously, remote impurity scattering rate decreases, and overall mobility remains almost constant for all dielectrics. 
We show the magnetic field dependence of the Hall mobility in Fig. \ref{Hall mobility vs magnetic field at T 300 K and 1e16 doping}. At room temperature, the Hall mobility is almost constant for a low magnetic field and decreases as we increase the magnetic field value for suspended MoS\textsubscript{2}. While for MoS\textsubscript{2} over the substrate, it remains almost constant.

\indent To demonstrate the tremendous potential of MoS\textsubscript{2} based sensors, the MR is an important figure of merit in such devices. In order to shed light on the relationship between the MR and magnetic field, we present an overview of the magnetic field dependence of the MR for a wide range of temperatures for a carrier density of $1\times 10^{12}~cm^{-2}$. The low-temperature MR factor in Fig. \ref{MR without substrate} is close to unity. At higher temperatures, the combined effect of ADP and POP scattering causes a reduction in the MR. In Fig. \ref{MR in substrate less MoS2}, we plotted the variation of MR with a magnetic field for suspended and other substrate materials. We demonstrated that, at low fields, MR increases with the magnetic field and then reaches saturation at higher fields. To read the magnetic data on a computer, such quantity can be helpful. For real-world applications, a high MR value at room temperature is needed.

\begin{figure}[!t]
	\centering	{\includegraphics[height=0.35\textwidth,width=0.45\textwidth]{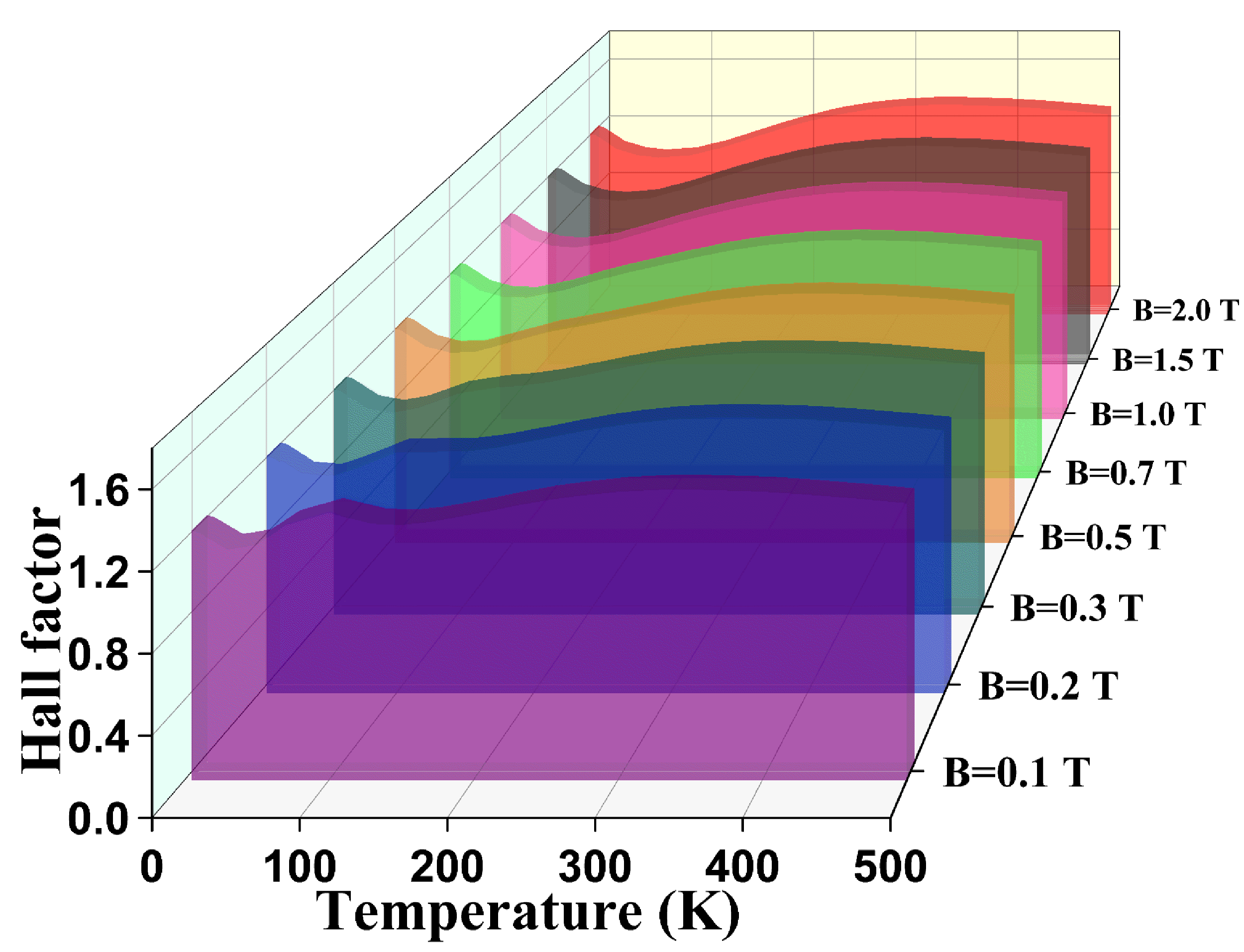}}
	\quad
	\caption{Temperature dependence of the Hall factor at various magnetic field values for a carrier density of $1\times 10^{12}~cm^{-2}$ in the suspended monolayer MoS\textsubscript{2}.}
	\label{Hall factor without substrate}
\end{figure}

\begin{figure}[!t]
	\centering
	\subfigure[]{\includegraphics[height=0.35\textwidth,width=0.45\textwidth]{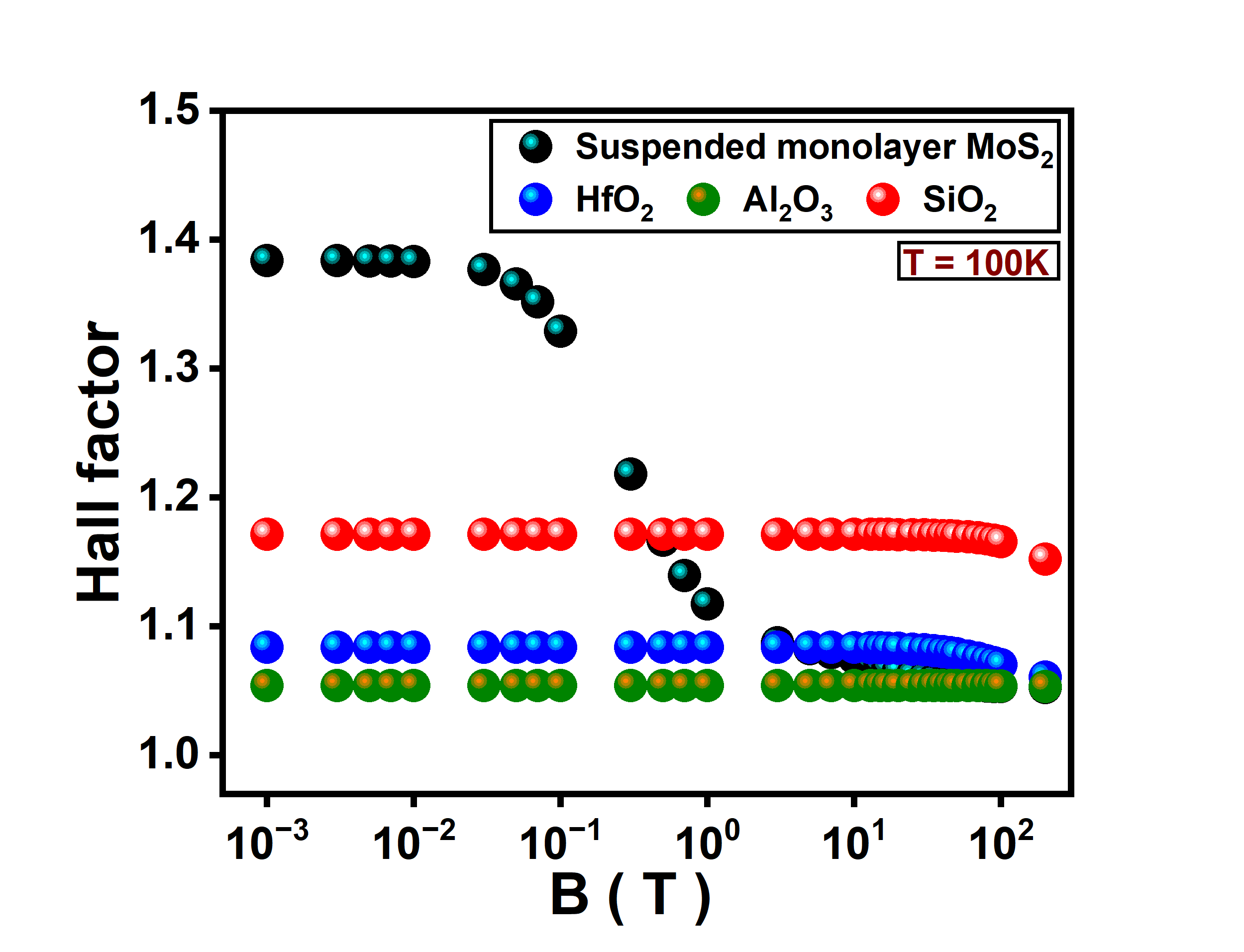}\label{Hall factor vs magnetic field at T 100K and 1e16 doping}}
		\quad
	\subfigure[]{\includegraphics[height=0.35\textwidth,width=0.45\textwidth]{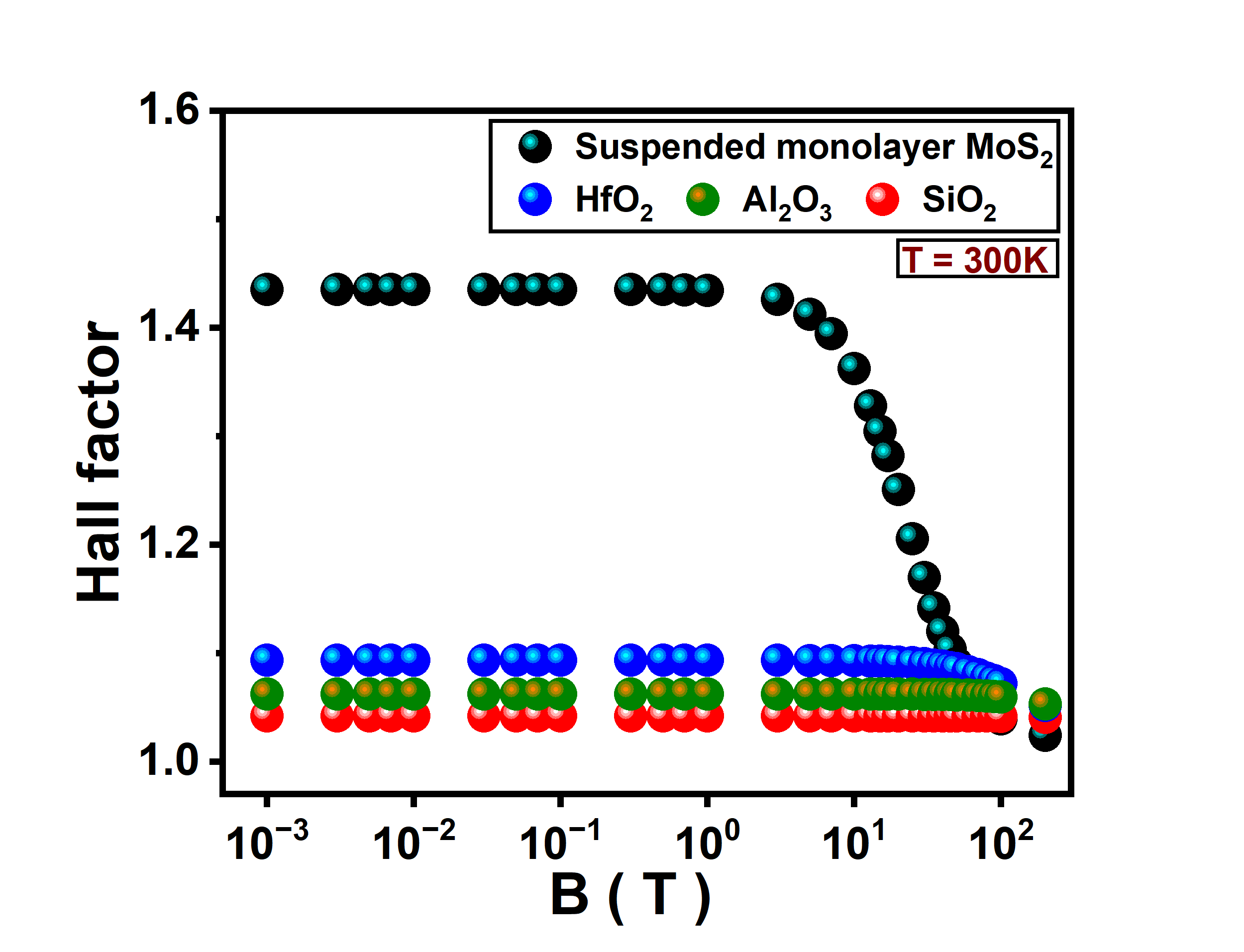}\label{Hall factor vs magnetic field at T 300K and 1e16 doping}}

\caption{The magnetic field dependence of the Hall factor (a) at T = 100 K and n = $1\times10^{12} ~ cm^{-2}$ and (b) T = 300 K and n = $1\times10^{12} ~ cm^{-2}$.}
\label{Hall factor vs magnetic field with and without substrate}
\end{figure}

The temperature-dependent behavior of the Hall scattering factor in a suspended monolayer MoS\textsubscript{2} is calculated at various B values as shown in Fig. \ref{Hall factor without substrate}. At higher temperatures compared to POP temperature $T_{POP}$, the Hall factor decreases to a lower value because POP scattering is weakly momentum dependent \cite{rode1975low}. At approximately half of the POP temperature $T_{POP}/2$, the Hall factor reaches its maximum value because POP scattering is a rapidly varying function at this energy. At a lower value of temperature around $60 $ K, the Hall factor falls to a lower value of around 1.1 because PZ, ADP, and POP scattering together gives a weakly momentum-dependent scattering rate \cite{rode1975low, rode1971electron}.

The magnetic field dependence of the Hall factors for monolayer MoS\textsubscript{2} at low and room temperature are plotted in Fig. \ref{Hall factor vs magnetic field with and without substrate}. The Hall factor of suspended MoS\textsubscript{2} have higher values of 1.38 and 1.43 for the lower magnetic field at 100 K and 300 K, respectively, and it remains constant for lower magnetic field and then starts decreasing at a higher magnetic field and reaches a lower value of around 1.05. In contrast, the Hall factor of MoS\textsubscript{2} over different substrates will be comparatively lower and does not vary much with the magnetic field. We demonstrated that at lower temperatures, the Hall factor of MoS\textsubscript{2} over Al\textsubscript{2}O\textsubscript{3} have a lower value of around 1.05 and remains almost constant for a whole range of magnetic field. 

\begin{figure}[!t]
	\centering
	\subfigure[]{\includegraphics[height=0.35\textwidth,width=0.45\textwidth]{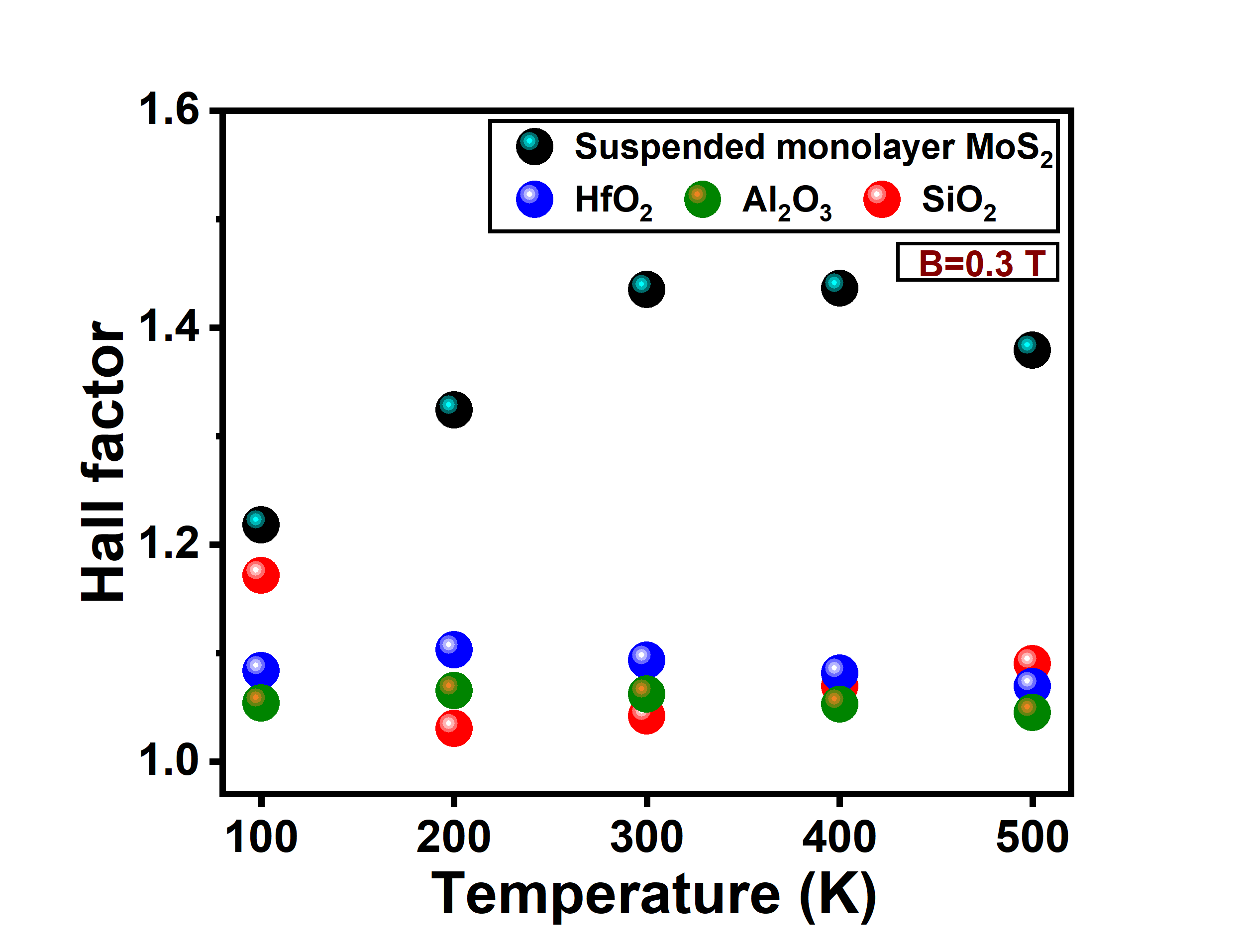}\label{Hall factor vs temp}}
		\quad
	\subfigure[]{\includegraphics[height=0.35\textwidth,width=0.45\textwidth]{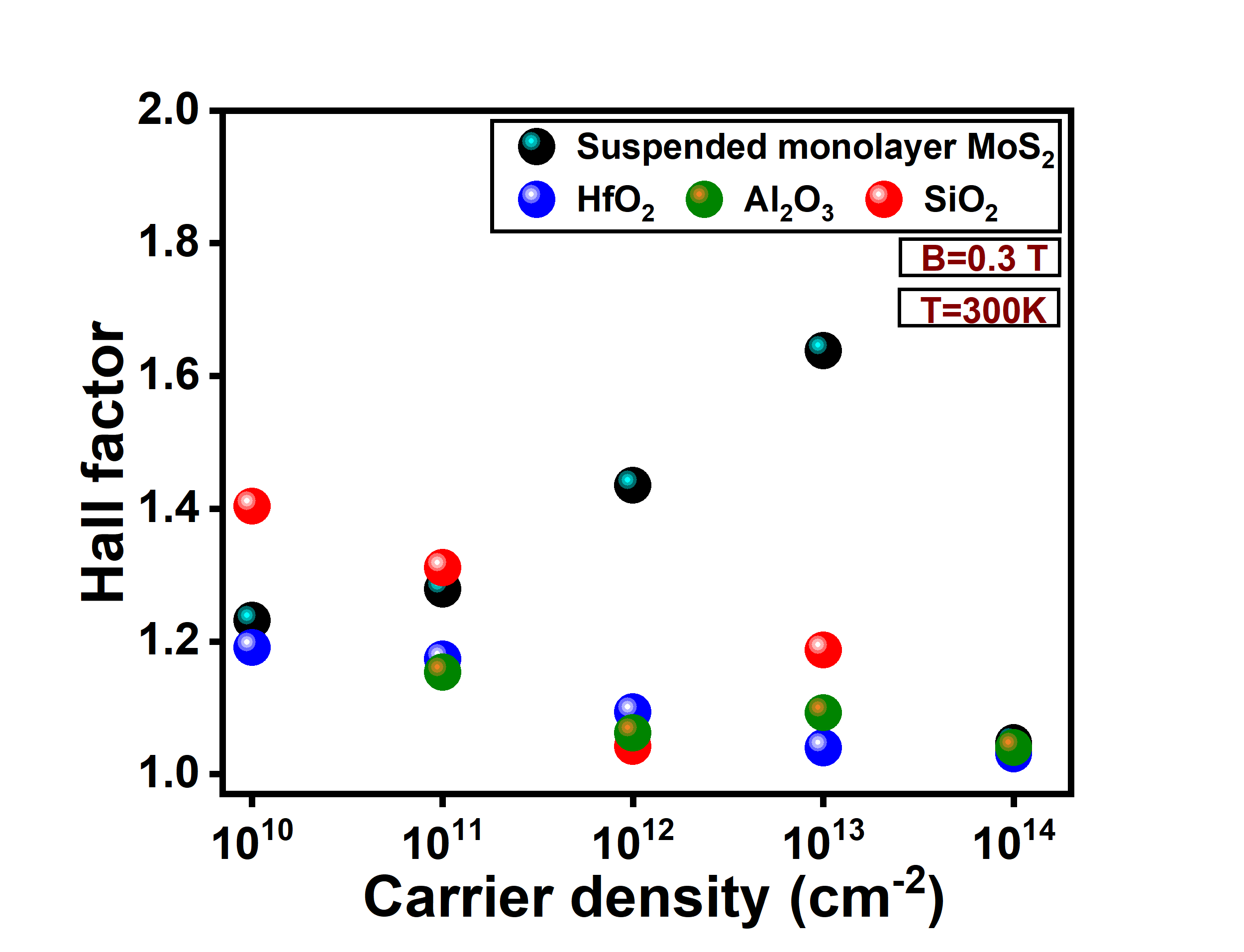}\label{Hall factor vs doping}}

\caption{The Hall factor in the monolayer MoS\textsubscript{2} at B = 0.3 T  (a) temperature dependence of the Hall factor at n = $1\times10^{12} ~ cm^{-2}$ (b) the Hall factor at room temperature as a function of the carrier density.}
\label{Hall factor vs temp and doping with and without substrate}
\end{figure}

\begin{figure}[!t]
	\centering
	\subfigure[]{\includegraphics[height=0.35\textwidth,width=0.45\textwidth]{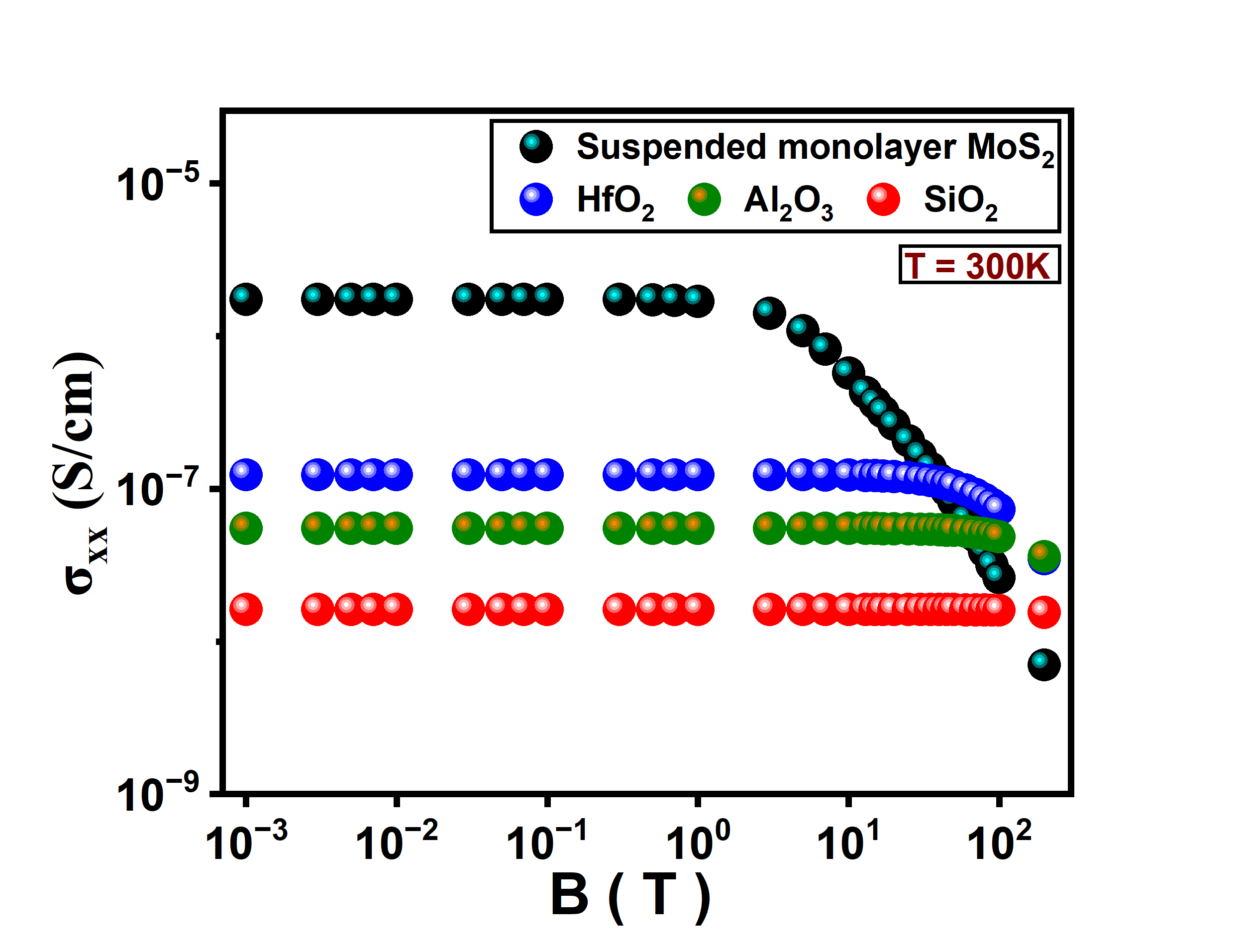}\label{Magnetoconductance vs magnetic field at T 300K and 1e16 doping }}
		\quad
	\subfigure[]{\includegraphics[height=0.35\textwidth,width=0.45\textwidth]{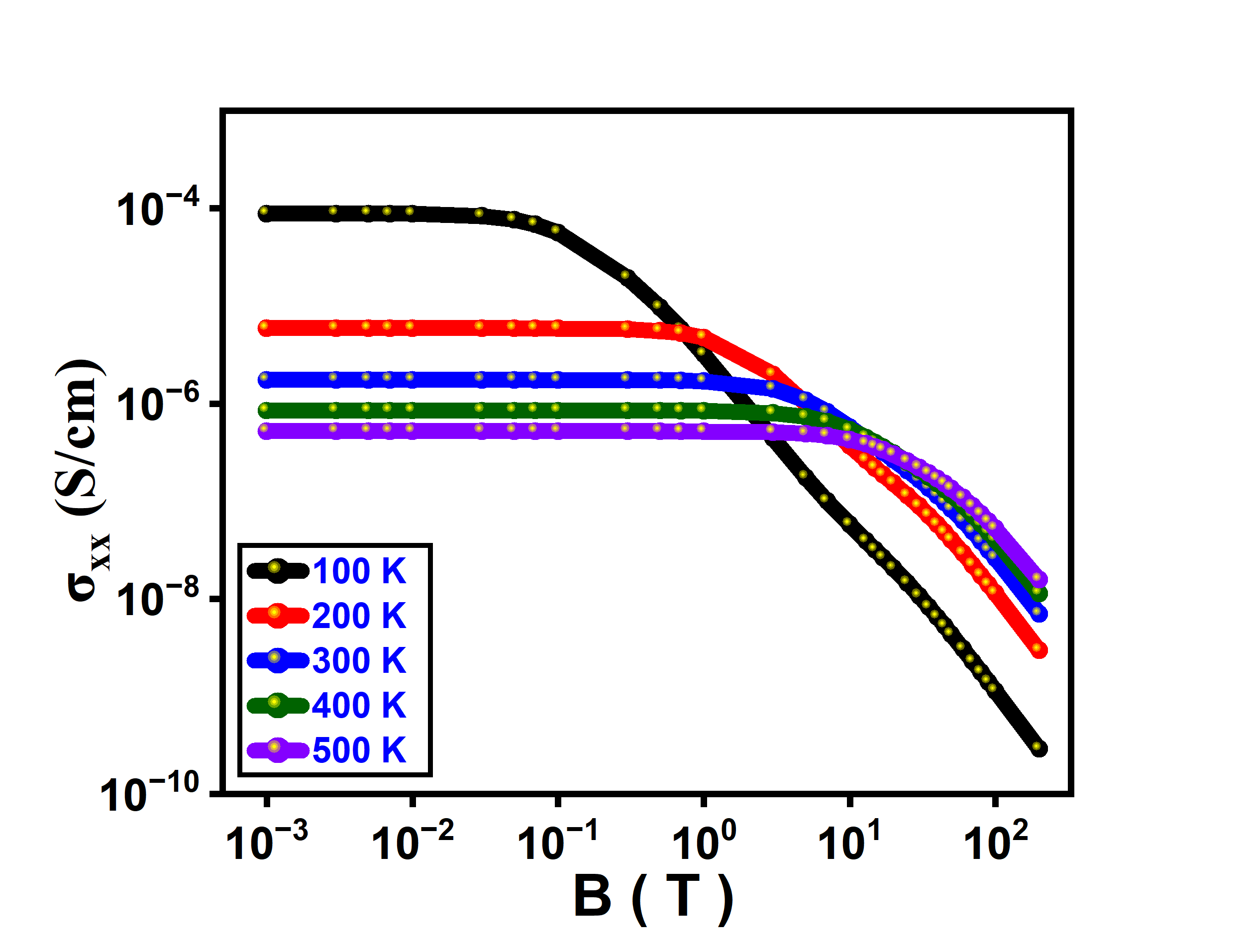}\label{sigma xx  vs magnetic field at 1e16 doping without substrate and various temp}}

\caption{Element of the conductivity tensor (a) the magnetic field dependence of the longitudinal conductivity at room temperature for a carrier density of $1\times 10^{12}~cm^{-2}$ (b) temperature-dependent longitudinal conductivity as a function of the magnetic field for a carrier density of $1\times 10^{12}~cm^{-2}$.}
\label{longitudinal_conductivity}
\end{figure}

To gain further insight into the electron transport in monolayer MoS\textsubscript{2}, we performed the Hall factor calculations for a wide range of temperatures and carrier densities. In Fig. \ref{Hall factor vs temp}, we showed the Hall factor variation with temperature for suspended monolayer MoS\textsubscript{2} and different substrate materials. The Hall factor for MoS\textsubscript{2} over different substrates has a lower value and does not vary much with temperature.  
In addition, we have calculated the Hall scattering factor for a variety of carrier densities as shown in Fig. \ref{Hall factor vs doping} and found that the Hall factor for MoS\textsubscript{2} with Al\textsubscript{2}O\textsubscript{3} substrate is a good choice of material for Hall effect detector since its Hall factor lies close to the ideal value of unity and remains stable for a wide range of magnetic field, temperature, and carrier concentration. 

To conclude our analysis, we plot the element of the conductivity tensor in Fig. \ref{longitudinal_conductivity} and study its behavior at various temperatures, focusing on their change due to the magnetic field for a constant carrier density. $\sigma_{xx}$ decreases as we increase the temperature. It decreases sharply at higher magnetic field strengths, whereas at lower magnetic field strengths, it is nearly constant. The decrease in conductivity at the higher magnetic field will be due to Lorentz's force deviating the charge carriers from their trajectory. As the temperature increases, the point of dramatic change also shifts towards a higher magnetic field.

\section{Conclusion}
\label{conclu}
In this work, we developed the electron transport model that accurately calculates the electron drift and Hall mobility, Hall factor, MR, and longitudinal conductivity in monolayer MoS\textsubscript{2}. 
It appears that corrections from Hall mobility to drift mobility for monolayer MoS\textsubscript{2} vary significantly depending on temperature, doping density, magnetic field strength, and dielectric material.\\
\indent In conclusion, the low and high-temperature transport parameters, such as Hall mobility and Hall scattering factor reported here, provide a platform for comparison with future measurement in monolayer MoS\textsubscript{2}, which can lead to experimental verification of the calculated parameters. It demonstrates the extent to which the Hall scattering factor can be affected by temperatures due to the polar and non-polar nature of phonons, which have turned out to be crucial for other TMDCs. Our results for monolayer MoS\textsubscript{2} leave room for improvements and call for stricter protocols for electrical and magneto-transport measurements in other TMDCs. The current formalism can be readily extended to investigate the MR and Hall scattering factor in other 2D materials and shed light on the origin of their unusual behavior of the MR and Hall factor. Our method is applicable to a wide variety of materials, such as semimetals, TMDCs, and many more.

\section*{Acknowledgments}
The authors gratefully acknowledge the funding from Indo-Korea Science and Technology Center (IKST), Bangalore. The author BM wishes to acknowledge the financial support from the Science and Engineering Research Board (SERB), Government of India, under Grant No. CRG/2021/003102 and Grant No. MTR/2021/000388.

\bibliography{reference}

\begin{thebibliography}{79}
\expandafter\ifx\csname natexlab\endcsname\relax\def\natexlab#1{#1}\fi
\expandafter\ifx\csname bibnamefont\endcsname\relax
  \def\bibnamefont#1{#1}\fi
\expandafter\ifx\csname bibfnamefont\endcsname\relax
  \def\bibfnamefont#1{#1}\fi
\expandafter\ifx\csname citenamefont\endcsname\relax
  \def\citenamefont#1{#1}\fi
\expandafter\ifx\csname url\endcsname\relax
  \def\url#1{\texttt{#1}}\fi
\expandafter\ifx\csname urlprefix\endcsname\relax\def\urlprefix{URL }\fi
\providecommand{\bibinfo}[2]{#2}
\providecommand{\eprint}[2][]{\url{#2}}

\bibitem[{\citenamefont{Mak and Shan}(2016)}]{mak2016photonics}
\bibinfo{author}{\bibfnamefont{K.~F.} \bibnamefont{Mak}} \bibnamefont{and}
  \bibinfo{author}{\bibfnamefont{J.}~\bibnamefont{Shan}},
  \bibinfo{journal}{Nature Photonics} \textbf{\bibinfo{volume}{10}},
  \bibinfo{pages}{216} (\bibinfo{year}{2016}).

\bibitem[{\citenamefont{Manzeli et~al.}(2017)\citenamefont{Manzeli,
  Ovchinnikov, Pasquier, Yazyev, and Kis}}]{manzeli20172d}
\bibinfo{author}{\bibfnamefont{S.}~\bibnamefont{Manzeli}},
  \bibinfo{author}{\bibfnamefont{D.}~\bibnamefont{Ovchinnikov}},
  \bibinfo{author}{\bibfnamefont{D.}~\bibnamefont{Pasquier}},
  \bibinfo{author}{\bibfnamefont{O.~V.} \bibnamefont{Yazyev}},
  \bibnamefont{and} \bibinfo{author}{\bibfnamefont{A.}~\bibnamefont{Kis}},
  \bibinfo{journal}{Nature Reviews Materials} \textbf{\bibinfo{volume}{2}},
  \bibinfo{pages}{1} (\bibinfo{year}{2017}).

\bibitem[{\citenamefont{Kolobov and Tominaga}(2016)}]{kolobov2016two}
\bibinfo{author}{\bibfnamefont{A.~V.} \bibnamefont{Kolobov}} \bibnamefont{and}
  \bibinfo{author}{\bibfnamefont{J.}~\bibnamefont{Tominaga}},
  \emph{\bibinfo{title}{Two-dimensional transition-metal dichalcogenides}},
  vol. \bibinfo{volume}{239} (\bibinfo{publisher}{Springer},
  \bibinfo{year}{2016}).

\bibitem[{\citenamefont{Chhowalla et~al.}(2015)\citenamefont{Chhowalla, Liu,
  and Zhang}}]{chhowalla2015two}
\bibinfo{author}{\bibfnamefont{M.}~\bibnamefont{Chhowalla}},
  \bibinfo{author}{\bibfnamefont{Z.}~\bibnamefont{Liu}}, \bibnamefont{and}
  \bibinfo{author}{\bibfnamefont{H.}~\bibnamefont{Zhang}},
  \bibinfo{journal}{Chemical Society Reviews} \textbf{\bibinfo{volume}{44}},
  \bibinfo{pages}{2584} (\bibinfo{year}{2015}).

\bibitem[{\citenamefont{Choi et~al.}(2017)\citenamefont{Choi, Choudhary, Han,
  Park, Akinwande, and Lee}}]{choi2017recent}
\bibinfo{author}{\bibfnamefont{W.}~\bibnamefont{Choi}},
  \bibinfo{author}{\bibfnamefont{N.}~\bibnamefont{Choudhary}},
  \bibinfo{author}{\bibfnamefont{G.~H.} \bibnamefont{Han}},
  \bibinfo{author}{\bibfnamefont{J.}~\bibnamefont{Park}},
  \bibinfo{author}{\bibfnamefont{D.}~\bibnamefont{Akinwande}},
  \bibnamefont{and} \bibinfo{author}{\bibfnamefont{Y.~H.} \bibnamefont{Lee}},
  \bibinfo{journal}{Materials Today} \textbf{\bibinfo{volume}{20}},
  \bibinfo{pages}{116} (\bibinfo{year}{2017}).

\bibitem[{\citenamefont{Yu et~al.}(2017)\citenamefont{Yu, Ong, Li, Xu, Zhang,
  Zhang, Shi, and Wang}}]{yu2017analyzing}
\bibinfo{author}{\bibfnamefont{Z.}~\bibnamefont{Yu}},
  \bibinfo{author}{\bibfnamefont{Z.-Y.} \bibnamefont{Ong}},
  \bibinfo{author}{\bibfnamefont{S.}~\bibnamefont{Li}},
  \bibinfo{author}{\bibfnamefont{J.-B.} \bibnamefont{Xu}},
  \bibinfo{author}{\bibfnamefont{G.}~\bibnamefont{Zhang}},
  \bibinfo{author}{\bibfnamefont{Y.-W.} \bibnamefont{Zhang}},
  \bibinfo{author}{\bibfnamefont{Y.}~\bibnamefont{Shi}}, \bibnamefont{and}
  \bibinfo{author}{\bibfnamefont{X.}~\bibnamefont{Wang}},
  \bibinfo{journal}{Advanced Functional Materials}
  \textbf{\bibinfo{volume}{27}}, \bibinfo{pages}{1604093}
  (\bibinfo{year}{2017}).

\bibitem[{\citenamefont{Wang et~al.}(2012)\citenamefont{Wang, Kalantar-Zadeh,
  Kis, Coleman, and Strano}}]{wang2012electronics}
\bibinfo{author}{\bibfnamefont{Q.~H.} \bibnamefont{Wang}},
  \bibinfo{author}{\bibfnamefont{K.}~\bibnamefont{Kalantar-Zadeh}},
  \bibinfo{author}{\bibfnamefont{A.}~\bibnamefont{Kis}},
  \bibinfo{author}{\bibfnamefont{J.~N.} \bibnamefont{Coleman}},
  \bibnamefont{and} \bibinfo{author}{\bibfnamefont{M.~S.}
  \bibnamefont{Strano}}, \bibinfo{journal}{Nature nanotechnology}
  \textbf{\bibinfo{volume}{7}}, \bibinfo{pages}{699} (\bibinfo{year}{2012}).

\bibitem[{\citenamefont{Wilson and Yoffe}(1969)}]{wilson1969transition}
\bibinfo{author}{\bibfnamefont{J.~A.} \bibnamefont{Wilson}} \bibnamefont{and}
  \bibinfo{author}{\bibfnamefont{A.}~\bibnamefont{Yoffe}},
  \bibinfo{journal}{Advances in Physics} \textbf{\bibinfo{volume}{18}},
  \bibinfo{pages}{193} (\bibinfo{year}{1969}).

\bibitem[{\citenamefont{Zibouche et~al.}(2014)\citenamefont{Zibouche, Kuc,
  Musfeldt, and Heine}}]{zibouche2014transition}
\bibinfo{author}{\bibfnamefont{N.}~\bibnamefont{Zibouche}},
  \bibinfo{author}{\bibfnamefont{A.}~\bibnamefont{Kuc}},
  \bibinfo{author}{\bibfnamefont{J.}~\bibnamefont{Musfeldt}}, \bibnamefont{and}
  \bibinfo{author}{\bibfnamefont{T.}~\bibnamefont{Heine}},
  \bibinfo{journal}{Annalen der Physik} \textbf{\bibinfo{volume}{526}},
  \bibinfo{pages}{395} (\bibinfo{year}{2014}).

\bibitem[{\citenamefont{Chuang et~al.}(2016)\citenamefont{Chuang, Chamlagain,
  Koehler, Perera, Yan, Mandrus, Tomanek, and Zhou}}]{chuang2016low}
\bibinfo{author}{\bibfnamefont{H.-J.} \bibnamefont{Chuang}},
  \bibinfo{author}{\bibfnamefont{B.}~\bibnamefont{Chamlagain}},
  \bibinfo{author}{\bibfnamefont{M.}~\bibnamefont{Koehler}},
  \bibinfo{author}{\bibfnamefont{M.~M.} \bibnamefont{Perera}},
  \bibinfo{author}{\bibfnamefont{J.}~\bibnamefont{Yan}},
  \bibinfo{author}{\bibfnamefont{D.}~\bibnamefont{Mandrus}},
  \bibinfo{author}{\bibfnamefont{D.}~\bibnamefont{Tomanek}}, \bibnamefont{and}
  \bibinfo{author}{\bibfnamefont{Z.}~\bibnamefont{Zhou}},
  \bibinfo{journal}{Nano letters} \textbf{\bibinfo{volume}{16}},
  \bibinfo{pages}{1896} (\bibinfo{year}{2016}).

\bibitem[{\citenamefont{Schmidt et~al.}(2015)\citenamefont{Schmidt,
  Giustiniano, and Eda}}]{schmidt2015electronic}
\bibinfo{author}{\bibfnamefont{H.}~\bibnamefont{Schmidt}},
  \bibinfo{author}{\bibfnamefont{F.}~\bibnamefont{Giustiniano}},
  \bibnamefont{and} \bibinfo{author}{\bibfnamefont{G.}~\bibnamefont{Eda}},
  \bibinfo{journal}{Chemical Society Reviews} \textbf{\bibinfo{volume}{44}},
  \bibinfo{pages}{7715} (\bibinfo{year}{2015}).

\bibitem[{\citenamefont{Liu et~al.}(2013)\citenamefont{Liu, Kang, Sarkar,
  Khatami, Jena, and Banerjee}}]{liu2013role}
\bibinfo{author}{\bibfnamefont{W.}~\bibnamefont{Liu}},
  \bibinfo{author}{\bibfnamefont{J.}~\bibnamefont{Kang}},
  \bibinfo{author}{\bibfnamefont{D.}~\bibnamefont{Sarkar}},
  \bibinfo{author}{\bibfnamefont{Y.}~\bibnamefont{Khatami}},
  \bibinfo{author}{\bibfnamefont{D.}~\bibnamefont{Jena}}, \bibnamefont{and}
  \bibinfo{author}{\bibfnamefont{K.}~\bibnamefont{Banerjee}},
  \bibinfo{journal}{Nano letters} \textbf{\bibinfo{volume}{13}},
  \bibinfo{pages}{1983} (\bibinfo{year}{2013}).

\bibitem[{\citenamefont{Nourbakhsh et~al.}(2016)\citenamefont{Nourbakhsh,
  Zubair, Sajjad, Tavakkoli~KG, Chen, Fang, Ling, Kong, Dresselhaus, Kaxiras
  et~al.}}]{nourbakhsh2016mos2}
\bibinfo{author}{\bibfnamefont{A.}~\bibnamefont{Nourbakhsh}},
  \bibinfo{author}{\bibfnamefont{A.}~\bibnamefont{Zubair}},
  \bibinfo{author}{\bibfnamefont{R.~N.} \bibnamefont{Sajjad}},
  \bibinfo{author}{\bibfnamefont{A.}~\bibnamefont{Tavakkoli~KG}},
  \bibinfo{author}{\bibfnamefont{W.}~\bibnamefont{Chen}},
  \bibinfo{author}{\bibfnamefont{S.}~\bibnamefont{Fang}},
  \bibinfo{author}{\bibfnamefont{X.}~\bibnamefont{Ling}},
  \bibinfo{author}{\bibfnamefont{J.}~\bibnamefont{Kong}},
  \bibinfo{author}{\bibfnamefont{M.~S.} \bibnamefont{Dresselhaus}},
  \bibinfo{author}{\bibfnamefont{E.}~\bibnamefont{Kaxiras}},
  \bibnamefont{et~al.}, \bibinfo{journal}{Nano letters}
  \textbf{\bibinfo{volume}{16}}, \bibinfo{pages}{7798} (\bibinfo{year}{2016}).

\bibitem[{\citenamefont{Late et~al.}(2012)\citenamefont{Late, Liu, Matte,
  Dravid, and Rao}}]{late2012hysteresis}
\bibinfo{author}{\bibfnamefont{D.~J.} \bibnamefont{Late}},
  \bibinfo{author}{\bibfnamefont{B.}~\bibnamefont{Liu}},
  \bibinfo{author}{\bibfnamefont{H.~R.} \bibnamefont{Matte}},
  \bibinfo{author}{\bibfnamefont{V.~P.} \bibnamefont{Dravid}},
  \bibnamefont{and} \bibinfo{author}{\bibfnamefont{C.}~\bibnamefont{Rao}},
  \bibinfo{journal}{ACS nano} \textbf{\bibinfo{volume}{6}},
  \bibinfo{pages}{5635} (\bibinfo{year}{2012}).

\bibitem[{\citenamefont{Radisavljevic et~al.}(2011)\citenamefont{Radisavljevic,
  Radenovic, Brivio, Giacometti, and Kis}}]{radisavljevic2011single}
\bibinfo{author}{\bibfnamefont{B.}~\bibnamefont{Radisavljevic}},
  \bibinfo{author}{\bibfnamefont{A.}~\bibnamefont{Radenovic}},
  \bibinfo{author}{\bibfnamefont{J.}~\bibnamefont{Brivio}},
  \bibinfo{author}{\bibfnamefont{V.}~\bibnamefont{Giacometti}},
  \bibnamefont{and} \bibinfo{author}{\bibfnamefont{A.}~\bibnamefont{Kis}},
  \bibinfo{journal}{Nature nanotechnology} \textbf{\bibinfo{volume}{6}},
  \bibinfo{pages}{147} (\bibinfo{year}{2011}).

\bibitem[{\citenamefont{Das et~al.}(2014)\citenamefont{Das, Prakash, Salazar,
  and Appenzeller}}]{das2014toward}
\bibinfo{author}{\bibfnamefont{S.}~\bibnamefont{Das}},
  \bibinfo{author}{\bibfnamefont{A.}~\bibnamefont{Prakash}},
  \bibinfo{author}{\bibfnamefont{R.}~\bibnamefont{Salazar}}, \bibnamefont{and}
  \bibinfo{author}{\bibfnamefont{J.}~\bibnamefont{Appenzeller}},
  \bibinfo{journal}{ACS nano} \textbf{\bibinfo{volume}{8}},
  \bibinfo{pages}{1681} (\bibinfo{year}{2014}).

\bibitem[{\citenamefont{Perera et~al.}(2013)\citenamefont{Perera, Lin, Chuang,
  Chamlagain, Wang, Tan, Cheng, Tom{\'a}nek, and Zhou}}]{perera2013improved}
\bibinfo{author}{\bibfnamefont{M.~M.} \bibnamefont{Perera}},
  \bibinfo{author}{\bibfnamefont{M.-W.} \bibnamefont{Lin}},
  \bibinfo{author}{\bibfnamefont{H.-J.} \bibnamefont{Chuang}},
  \bibinfo{author}{\bibfnamefont{B.~P.} \bibnamefont{Chamlagain}},
  \bibinfo{author}{\bibfnamefont{C.}~\bibnamefont{Wang}},
  \bibinfo{author}{\bibfnamefont{X.}~\bibnamefont{Tan}},
  \bibinfo{author}{\bibfnamefont{M.~M.-C.} \bibnamefont{Cheng}},
  \bibinfo{author}{\bibfnamefont{D.}~\bibnamefont{Tom{\'a}nek}},
  \bibnamefont{and} \bibinfo{author}{\bibfnamefont{Z.}~\bibnamefont{Zhou}},
  \bibinfo{journal}{ACS nano} \textbf{\bibinfo{volume}{7}},
  \bibinfo{pages}{4449} (\bibinfo{year}{2013}).

\bibitem[{\citenamefont{Bao et~al.}(2013)\citenamefont{Bao, Cai, Kim, Sridhara,
  and Fuhrer}}]{bao2013high}
\bibinfo{author}{\bibfnamefont{W.}~\bibnamefont{Bao}},
  \bibinfo{author}{\bibfnamefont{X.}~\bibnamefont{Cai}},
  \bibinfo{author}{\bibfnamefont{D.}~\bibnamefont{Kim}},
  \bibinfo{author}{\bibfnamefont{K.}~\bibnamefont{Sridhara}}, \bibnamefont{and}
  \bibinfo{author}{\bibfnamefont{M.~S.} \bibnamefont{Fuhrer}},
  \bibinfo{journal}{Applied Physics Letters} \textbf{\bibinfo{volume}{102}},
  \bibinfo{pages}{042104} (\bibinfo{year}{2013}).

\bibitem[{\citenamefont{Illarionov et~al.}(2018)\citenamefont{Illarionov,
  Smithe, Waltl, Grady, Deshmukh, Pop, and Grasser}}]{illarionov2018annealing}
\bibinfo{author}{\bibfnamefont{Y.~Y.} \bibnamefont{Illarionov}},
  \bibinfo{author}{\bibfnamefont{K.~K.} \bibnamefont{Smithe}},
  \bibinfo{author}{\bibfnamefont{M.}~\bibnamefont{Waltl}},
  \bibinfo{author}{\bibfnamefont{R.~W.} \bibnamefont{Grady}},
  \bibinfo{author}{\bibfnamefont{S.}~\bibnamefont{Deshmukh}},
  \bibinfo{author}{\bibfnamefont{E.}~\bibnamefont{Pop}}, \bibnamefont{and}
  \bibinfo{author}{\bibfnamefont{T.}~\bibnamefont{Grasser}}, in
  \emph{\bibinfo{booktitle}{2018 76th Device Research Conference (DRC)}}
  (\bibinfo{organization}{IEEE}, \bibinfo{year}{2018}), pp.
  \bibinfo{pages}{1--2}.

\bibitem[{\citenamefont{Zhang et~al.}(2015)\citenamefont{Zhang, Xie, Xu, Sun,
  Li, Zhang, Dai, Zhao, Li, Li et~al.}}]{zhang2015mos}
\bibinfo{author}{\bibfnamefont{X.-W.} \bibnamefont{Zhang}},
  \bibinfo{author}{\bibfnamefont{D.}~\bibnamefont{Xie}},
  \bibinfo{author}{\bibfnamefont{J.-L.} \bibnamefont{Xu}},
  \bibinfo{author}{\bibfnamefont{Y.-L.} \bibnamefont{Sun}},
  \bibinfo{author}{\bibfnamefont{X.}~\bibnamefont{Li}},
  \bibinfo{author}{\bibfnamefont{C.}~\bibnamefont{Zhang}},
  \bibinfo{author}{\bibfnamefont{R.-X.} \bibnamefont{Dai}},
  \bibinfo{author}{\bibfnamefont{Y.-F.} \bibnamefont{Zhao}},
  \bibinfo{author}{\bibfnamefont{X.-M.} \bibnamefont{Li}},
  \bibinfo{author}{\bibfnamefont{X.}~\bibnamefont{Li}}, \bibnamefont{et~al.},
  \bibinfo{journal}{IEEE Electron Device Letters}
  \textbf{\bibinfo{volume}{36}}, \bibinfo{pages}{784} (\bibinfo{year}{2015}).

\bibitem[{\citenamefont{Li et~al.}(2017)\citenamefont{Li, Yan, Bao, Ding,
  Zhang, and Zhou}}]{li2017low}
\bibinfo{author}{\bibfnamefont{C.}~\bibnamefont{Li}},
  \bibinfo{author}{\bibfnamefont{X.}~\bibnamefont{Yan}},
  \bibinfo{author}{\bibfnamefont{W.}~\bibnamefont{Bao}},
  \bibinfo{author}{\bibfnamefont{S.}~\bibnamefont{Ding}},
  \bibinfo{author}{\bibfnamefont{D.~W.} \bibnamefont{Zhang}}, \bibnamefont{and}
  \bibinfo{author}{\bibfnamefont{P.}~\bibnamefont{Zhou}},
  \bibinfo{journal}{Applied Physics Letters} \textbf{\bibinfo{volume}{111}},
  \bibinfo{pages}{193502} (\bibinfo{year}{2017}).

\bibitem[{\citenamefont{Liu et~al.}(2018)\citenamefont{Liu, Liang, Gao, Pan,
  Jiang, Xu, Luo, Zou, Yang, Liao et~al.}}]{liu2018mos2}
\bibinfo{author}{\bibfnamefont{X.}~\bibnamefont{Liu}},
  \bibinfo{author}{\bibfnamefont{R.}~\bibnamefont{Liang}},
  \bibinfo{author}{\bibfnamefont{G.}~\bibnamefont{Gao}},
  \bibinfo{author}{\bibfnamefont{C.}~\bibnamefont{Pan}},
  \bibinfo{author}{\bibfnamefont{C.}~\bibnamefont{Jiang}},
  \bibinfo{author}{\bibfnamefont{Q.}~\bibnamefont{Xu}},
  \bibinfo{author}{\bibfnamefont{J.}~\bibnamefont{Luo}},
  \bibinfo{author}{\bibfnamefont{X.}~\bibnamefont{Zou}},
  \bibinfo{author}{\bibfnamefont{Z.}~\bibnamefont{Yang}},
  \bibinfo{author}{\bibfnamefont{L.}~\bibnamefont{Liao}}, \bibnamefont{et~al.},
  \bibinfo{journal}{Advanced Materials} \textbf{\bibinfo{volume}{30}},
  \bibinfo{pages}{1800932} (\bibinfo{year}{2018}).

\bibitem[{\citenamefont{Iqbal et~al.}(2015)\citenamefont{Iqbal, Iqbal, Khan,
  Shehzad, Seo, Park, Hwang, and Eom}}]{iqbal2015high}
\bibinfo{author}{\bibfnamefont{M.~W.} \bibnamefont{Iqbal}},
  \bibinfo{author}{\bibfnamefont{M.~Z.} \bibnamefont{Iqbal}},
  \bibinfo{author}{\bibfnamefont{M.~F.} \bibnamefont{Khan}},
  \bibinfo{author}{\bibfnamefont{M.~A.} \bibnamefont{Shehzad}},
  \bibinfo{author}{\bibfnamefont{Y.}~\bibnamefont{Seo}},
  \bibinfo{author}{\bibfnamefont{J.~H.} \bibnamefont{Park}},
  \bibinfo{author}{\bibfnamefont{C.}~\bibnamefont{Hwang}}, \bibnamefont{and}
  \bibinfo{author}{\bibfnamefont{J.}~\bibnamefont{Eom}},
  \bibinfo{journal}{Scientific reports} \textbf{\bibinfo{volume}{5}},
  \bibinfo{pages}{10699} (\bibinfo{year}{2015}).

\bibitem[{\citenamefont{Sik~Hwang et~al.}(2012)\citenamefont{Sik~Hwang,
  Remskar, Yan, Protasenko, Tahy, Doo~Chae, Zhao, Konar, Xing, Seabaugh
  et~al.}}]{sik2012transistors}
\bibinfo{author}{\bibfnamefont{W.}~\bibnamefont{Sik~Hwang}},
  \bibinfo{author}{\bibfnamefont{M.}~\bibnamefont{Remskar}},
  \bibinfo{author}{\bibfnamefont{R.}~\bibnamefont{Yan}},
  \bibinfo{author}{\bibfnamefont{V.}~\bibnamefont{Protasenko}},
  \bibinfo{author}{\bibfnamefont{K.}~\bibnamefont{Tahy}},
  \bibinfo{author}{\bibfnamefont{S.}~\bibnamefont{Doo~Chae}},
  \bibinfo{author}{\bibfnamefont{P.}~\bibnamefont{Zhao}},
  \bibinfo{author}{\bibfnamefont{A.}~\bibnamefont{Konar}},
  \bibinfo{author}{\bibfnamefont{H.}~\bibnamefont{Xing}},
  \bibinfo{author}{\bibfnamefont{A.}~\bibnamefont{Seabaugh}},
  \bibnamefont{et~al.}, \bibinfo{journal}{Applied physics letters}
  \textbf{\bibinfo{volume}{101}}, \bibinfo{pages}{013107}
  (\bibinfo{year}{2012}).

\bibitem[{\citenamefont{Zhang et~al.}(2012)\citenamefont{Zhang, Ye, Matsuhashi,
  and Iwasa}}]{zhang2012ambipolar}
\bibinfo{author}{\bibfnamefont{Y.}~\bibnamefont{Zhang}},
  \bibinfo{author}{\bibfnamefont{J.}~\bibnamefont{Ye}},
  \bibinfo{author}{\bibfnamefont{Y.}~\bibnamefont{Matsuhashi}},
  \bibnamefont{and} \bibinfo{author}{\bibfnamefont{Y.}~\bibnamefont{Iwasa}},
  \bibinfo{journal}{Nano letters} \textbf{\bibinfo{volume}{12}},
  \bibinfo{pages}{1136} (\bibinfo{year}{2012}).

\bibitem[{\citenamefont{Lopez-Sanchez et~al.}(2013)\citenamefont{Lopez-Sanchez,
  Lembke, Kayci, Radenovic, and Kis}}]{lopez2013ultrasensitive}
\bibinfo{author}{\bibfnamefont{O.}~\bibnamefont{Lopez-Sanchez}},
  \bibinfo{author}{\bibfnamefont{D.}~\bibnamefont{Lembke}},
  \bibinfo{author}{\bibfnamefont{M.}~\bibnamefont{Kayci}},
  \bibinfo{author}{\bibfnamefont{A.}~\bibnamefont{Radenovic}},
  \bibnamefont{and} \bibinfo{author}{\bibfnamefont{A.}~\bibnamefont{Kis}},
  \bibinfo{journal}{Nature nanotechnology} \textbf{\bibinfo{volume}{8}},
  \bibinfo{pages}{497} (\bibinfo{year}{2013}).

\bibitem[{\citenamefont{Lembke et~al.}(2015)\citenamefont{Lembke, Bertolazzi,
  and Kis}}]{lembke2015single}
\bibinfo{author}{\bibfnamefont{D.}~\bibnamefont{Lembke}},
  \bibinfo{author}{\bibfnamefont{S.}~\bibnamefont{Bertolazzi}},
  \bibnamefont{and} \bibinfo{author}{\bibfnamefont{A.}~\bibnamefont{Kis}},
  \bibinfo{journal}{Accounts of chemical research}
  \textbf{\bibinfo{volume}{48}}, \bibinfo{pages}{100} (\bibinfo{year}{2015}).

\bibitem[{\citenamefont{Pham et~al.}(2019)\citenamefont{Pham, Li, Bekyarova,
  Itkis, and Mulchandani}}]{pham2019mos2}
\bibinfo{author}{\bibfnamefont{T.}~\bibnamefont{Pham}},
  \bibinfo{author}{\bibfnamefont{G.}~\bibnamefont{Li}},
  \bibinfo{author}{\bibfnamefont{E.}~\bibnamefont{Bekyarova}},
  \bibinfo{author}{\bibfnamefont{M.~E.} \bibnamefont{Itkis}}, \bibnamefont{and}
  \bibinfo{author}{\bibfnamefont{A.}~\bibnamefont{Mulchandani}},
  \bibinfo{journal}{ACS nano} \textbf{\bibinfo{volume}{13}},
  \bibinfo{pages}{3196} (\bibinfo{year}{2019}).

\bibitem[{\citenamefont{Singh et~al.}(2019)\citenamefont{Singh, Singh, Kim,
  Yeom, and Nalwa}}]{singh2019flexible}
\bibinfo{author}{\bibfnamefont{E.}~\bibnamefont{Singh}},
  \bibinfo{author}{\bibfnamefont{P.}~\bibnamefont{Singh}},
  \bibinfo{author}{\bibfnamefont{K.~S.} \bibnamefont{Kim}},
  \bibinfo{author}{\bibfnamefont{G.~Y.} \bibnamefont{Yeom}}, \bibnamefont{and}
  \bibinfo{author}{\bibfnamefont{H.~S.} \bibnamefont{Nalwa}},
  \bibinfo{journal}{ACS applied materials \& interfaces}
  \textbf{\bibinfo{volume}{11}}, \bibinfo{pages}{11061} (\bibinfo{year}{2019}).

\bibitem[{\citenamefont{Choi et~al.}(2014)\citenamefont{Choi, Qu, Lee, Liu,
  Watanabe, Taniguchi, and Yoo}}]{choi2014lateral}
\bibinfo{author}{\bibfnamefont{M.~S.} \bibnamefont{Choi}},
  \bibinfo{author}{\bibfnamefont{D.}~\bibnamefont{Qu}},
  \bibinfo{author}{\bibfnamefont{D.}~\bibnamefont{Lee}},
  \bibinfo{author}{\bibfnamefont{X.}~\bibnamefont{Liu}},
  \bibinfo{author}{\bibfnamefont{K.}~\bibnamefont{Watanabe}},
  \bibinfo{author}{\bibfnamefont{T.}~\bibnamefont{Taniguchi}},
  \bibnamefont{and} \bibinfo{author}{\bibfnamefont{W.~J.} \bibnamefont{Yoo}},
  \bibinfo{journal}{ACS nano} \textbf{\bibinfo{volume}{8}},
  \bibinfo{pages}{9332} (\bibinfo{year}{2014}).

\bibitem[{\citenamefont{Voiry et~al.}(2013)\citenamefont{Voiry, Salehi, Silva,
  Fujita, Chen, Asefa, Shenoy, Eda, and Chhowalla}}]{voiry2013conducting}
\bibinfo{author}{\bibfnamefont{D.}~\bibnamefont{Voiry}},
  \bibinfo{author}{\bibfnamefont{M.}~\bibnamefont{Salehi}},
  \bibinfo{author}{\bibfnamefont{R.}~\bibnamefont{Silva}},
  \bibinfo{author}{\bibfnamefont{T.}~\bibnamefont{Fujita}},
  \bibinfo{author}{\bibfnamefont{M.}~\bibnamefont{Chen}},
  \bibinfo{author}{\bibfnamefont{T.}~\bibnamefont{Asefa}},
  \bibinfo{author}{\bibfnamefont{V.~B.} \bibnamefont{Shenoy}},
  \bibinfo{author}{\bibfnamefont{G.}~\bibnamefont{Eda}}, \bibnamefont{and}
  \bibinfo{author}{\bibfnamefont{M.}~\bibnamefont{Chhowalla}},
  \bibinfo{journal}{Nano letters} \textbf{\bibinfo{volume}{13}},
  \bibinfo{pages}{6222} (\bibinfo{year}{2013}).

\bibitem[{\citenamefont{Gali et~al.}(2020)\citenamefont{Gali, Pershin,
  Lherbier, Charlier, and Beljonne}}]{gali2020electronic}
\bibinfo{author}{\bibfnamefont{S.~M.} \bibnamefont{Gali}},
  \bibinfo{author}{\bibfnamefont{A.}~\bibnamefont{Pershin}},
  \bibinfo{author}{\bibfnamefont{A.}~\bibnamefont{Lherbier}},
  \bibinfo{author}{\bibfnamefont{J.-C.} \bibnamefont{Charlier}},
  \bibnamefont{and} \bibinfo{author}{\bibfnamefont{D.}~\bibnamefont{Beljonne}},
  \bibinfo{journal}{The Journal of Physical Chemistry C}
  \textbf{\bibinfo{volume}{124}}, \bibinfo{pages}{15076}
  (\bibinfo{year}{2020}).

\bibitem[{\citenamefont{Lukowski et~al.}(2013)\citenamefont{Lukowski, Daniel,
  Meng, Forticaux, Li, and Jin}}]{lukowski2013enhanced}
\bibinfo{author}{\bibfnamefont{M.~A.} \bibnamefont{Lukowski}},
  \bibinfo{author}{\bibfnamefont{A.~S.} \bibnamefont{Daniel}},
  \bibinfo{author}{\bibfnamefont{F.}~\bibnamefont{Meng}},
  \bibinfo{author}{\bibfnamefont{A.}~\bibnamefont{Forticaux}},
  \bibinfo{author}{\bibfnamefont{L.}~\bibnamefont{Li}}, \bibnamefont{and}
  \bibinfo{author}{\bibfnamefont{S.}~\bibnamefont{Jin}},
  \bibinfo{journal}{Journal of the American Chemical Society}
  \textbf{\bibinfo{volume}{135}}, \bibinfo{pages}{10274}
  (\bibinfo{year}{2013}).

\bibitem[{\citenamefont{Liu et~al.}(2012)\citenamefont{Liu, Neal, and
  Ye}}]{liu2012channel}
\bibinfo{author}{\bibfnamefont{H.}~\bibnamefont{Liu}},
  \bibinfo{author}{\bibfnamefont{A.~T.} \bibnamefont{Neal}}, \bibnamefont{and}
  \bibinfo{author}{\bibfnamefont{P.~D.} \bibnamefont{Ye}},
  \bibinfo{journal}{ACS nano} \textbf{\bibinfo{volume}{6}},
  \bibinfo{pages}{8563} (\bibinfo{year}{2012}).

\bibitem[{\citenamefont{Ghatak and Ghosh}(2013)}]{ghatak2013observation}
\bibinfo{author}{\bibfnamefont{S.}~\bibnamefont{Ghatak}} \bibnamefont{and}
  \bibinfo{author}{\bibfnamefont{A.}~\bibnamefont{Ghosh}},
  \bibinfo{journal}{Applied Physics Letters} \textbf{\bibinfo{volume}{103}},
  \bibinfo{pages}{122103} (\bibinfo{year}{2013}).

\bibitem[{\citenamefont{Li et~al.}(2014{\natexlab{a}})\citenamefont{Li,
  Komatsu, Nakaharai, Lin, Yamamoto, Duan, and Tsukagoshi}}]{li2014thickness}
\bibinfo{author}{\bibfnamefont{S.-L.} \bibnamefont{Li}},
  \bibinfo{author}{\bibfnamefont{K.}~\bibnamefont{Komatsu}},
  \bibinfo{author}{\bibfnamefont{S.}~\bibnamefont{Nakaharai}},
  \bibinfo{author}{\bibfnamefont{Y.-F.} \bibnamefont{Lin}},
  \bibinfo{author}{\bibfnamefont{M.}~\bibnamefont{Yamamoto}},
  \bibinfo{author}{\bibfnamefont{X.}~\bibnamefont{Duan}}, \bibnamefont{and}
  \bibinfo{author}{\bibfnamefont{K.}~\bibnamefont{Tsukagoshi}},
  \bibinfo{journal}{ACS nano} \textbf{\bibinfo{volume}{8}},
  \bibinfo{pages}{12836} (\bibinfo{year}{2014}{\natexlab{a}}).

\bibitem[{\citenamefont{Fei et~al.}(2016)\citenamefont{Fei, Lei, Zhang, Lu,
  Lin, Lam, Chai, and Wang}}]{fei2016direct}
\bibinfo{author}{\bibfnamefont{L.}~\bibnamefont{Fei}},
  \bibinfo{author}{\bibfnamefont{S.}~\bibnamefont{Lei}},
  \bibinfo{author}{\bibfnamefont{W.-B.} \bibnamefont{Zhang}},
  \bibinfo{author}{\bibfnamefont{W.}~\bibnamefont{Lu}},
  \bibinfo{author}{\bibfnamefont{Z.}~\bibnamefont{Lin}},
  \bibinfo{author}{\bibfnamefont{C.~H.} \bibnamefont{Lam}},
  \bibinfo{author}{\bibfnamefont{Y.}~\bibnamefont{Chai}}, \bibnamefont{and}
  \bibinfo{author}{\bibfnamefont{Y.}~\bibnamefont{Wang}},
  \bibinfo{journal}{Nature communications} \textbf{\bibinfo{volume}{7}},
  \bibinfo{pages}{12206} (\bibinfo{year}{2016}).

\bibitem[{\citenamefont{Dhyani and Das}(2017)}]{dhyani2017high}
\bibinfo{author}{\bibfnamefont{V.}~\bibnamefont{Dhyani}} \bibnamefont{and}
  \bibinfo{author}{\bibfnamefont{S.}~\bibnamefont{Das}},
  \bibinfo{journal}{Scientific reports} \textbf{\bibinfo{volume}{7}},
  \bibinfo{pages}{44243} (\bibinfo{year}{2017}).

\bibitem[{\citenamefont{Li et~al.}(2016{\natexlab{a}})\citenamefont{Li, Chen,
  Dhall, and Cronin}}]{li2016highly}
\bibinfo{author}{\bibfnamefont{Z.}~\bibnamefont{Li}},
  \bibinfo{author}{\bibfnamefont{J.}~\bibnamefont{Chen}},
  \bibinfo{author}{\bibfnamefont{R.}~\bibnamefont{Dhall}}, \bibnamefont{and}
  \bibinfo{author}{\bibfnamefont{S.~B.} \bibnamefont{Cronin}},
  \bibinfo{journal}{2D Materials} \textbf{\bibinfo{volume}{4}},
  \bibinfo{pages}{015004} (\bibinfo{year}{2016}{\natexlab{a}}).

\bibitem[{\citenamefont{Kufer and Konstantatos}(2015)}]{kufer2015highly}
\bibinfo{author}{\bibfnamefont{D.}~\bibnamefont{Kufer}} \bibnamefont{and}
  \bibinfo{author}{\bibfnamefont{G.}~\bibnamefont{Konstantatos}},
  \bibinfo{journal}{Nano letters} \textbf{\bibinfo{volume}{15}},
  \bibinfo{pages}{7307} (\bibinfo{year}{2015}).

\bibitem[{\citenamefont{Zhao et~al.}(2017)\citenamefont{Zhao, Li, Yu, Wei,
  Liao, Chen, Wang, Shi, Sun, and Zhang}}]{zhao2017highly}
\bibinfo{author}{\bibfnamefont{J.}~\bibnamefont{Zhao}},
  \bibinfo{author}{\bibfnamefont{N.}~\bibnamefont{Li}},
  \bibinfo{author}{\bibfnamefont{H.}~\bibnamefont{Yu}},
  \bibinfo{author}{\bibfnamefont{Z.}~\bibnamefont{Wei}},
  \bibinfo{author}{\bibfnamefont{M.}~\bibnamefont{Liao}},
  \bibinfo{author}{\bibfnamefont{P.}~\bibnamefont{Chen}},
  \bibinfo{author}{\bibfnamefont{S.}~\bibnamefont{Wang}},
  \bibinfo{author}{\bibfnamefont{D.}~\bibnamefont{Shi}},
  \bibinfo{author}{\bibfnamefont{Q.}~\bibnamefont{Sun}}, \bibnamefont{and}
  \bibinfo{author}{\bibfnamefont{G.}~\bibnamefont{Zhang}},
  \bibinfo{journal}{Advanced materials} \textbf{\bibinfo{volume}{29}},
  \bibinfo{pages}{1702076} (\bibinfo{year}{2017}).

\bibitem[{\citenamefont{Wang et~al.}(2022)\citenamefont{Wang, Song, and
  Huang}}]{wang2022evolution}
\bibinfo{author}{\bibfnamefont{C.}~\bibnamefont{Wang}},
  \bibinfo{author}{\bibfnamefont{Y.}~\bibnamefont{Song}}, \bibnamefont{and}
  \bibinfo{author}{\bibfnamefont{H.}~\bibnamefont{Huang}},
  \bibinfo{journal}{Nanomaterials} \textbf{\bibinfo{volume}{12}},
  \bibinfo{pages}{3233} (\bibinfo{year}{2022}).

\bibitem[{\citenamefont{Mak et~al.}(2013)\citenamefont{Mak, He, Lee, Lee, Hone,
  Heinz, and Shan}}]{mak2013tightly}
\bibinfo{author}{\bibfnamefont{K.~F.} \bibnamefont{Mak}},
  \bibinfo{author}{\bibfnamefont{K.}~\bibnamefont{He}},
  \bibinfo{author}{\bibfnamefont{C.}~\bibnamefont{Lee}},
  \bibinfo{author}{\bibfnamefont{G.~H.} \bibnamefont{Lee}},
  \bibinfo{author}{\bibfnamefont{J.}~\bibnamefont{Hone}},
  \bibinfo{author}{\bibfnamefont{T.~F.} \bibnamefont{Heinz}}, \bibnamefont{and}
  \bibinfo{author}{\bibfnamefont{J.}~\bibnamefont{Shan}},
  \bibinfo{journal}{Nature materials} \textbf{\bibinfo{volume}{12}},
  \bibinfo{pages}{207} (\bibinfo{year}{2013}).

\bibitem[{\citenamefont{Mak et~al.}(2010)\citenamefont{Mak, Lee, Hone, Shan,
  and Heinz}}]{mak2010atomically}
\bibinfo{author}{\bibfnamefont{K.~F.} \bibnamefont{Mak}},
  \bibinfo{author}{\bibfnamefont{C.}~\bibnamefont{Lee}},
  \bibinfo{author}{\bibfnamefont{J.}~\bibnamefont{Hone}},
  \bibinfo{author}{\bibfnamefont{J.}~\bibnamefont{Shan}}, \bibnamefont{and}
  \bibinfo{author}{\bibfnamefont{T.~F.} \bibnamefont{Heinz}},
  \bibinfo{journal}{Physical review letters} \textbf{\bibinfo{volume}{105}},
  \bibinfo{pages}{136805} (\bibinfo{year}{2010}).

\bibitem[{\citenamefont{Kaasbjerg et~al.}(2013)\citenamefont{Kaasbjerg,
  Thygesen, and Jauho}}]{kaasbjerg2013acoustic}
\bibinfo{author}{\bibfnamefont{K.}~\bibnamefont{Kaasbjerg}},
  \bibinfo{author}{\bibfnamefont{K.~S.} \bibnamefont{Thygesen}},
  \bibnamefont{and} \bibinfo{author}{\bibfnamefont{A.-P.} \bibnamefont{Jauho}},
  \bibinfo{journal}{Physical Review B} \textbf{\bibinfo{volume}{87}},
  \bibinfo{pages}{235312} (\bibinfo{year}{2013}).

\bibitem[{\citenamefont{Ganatra and Zhang}(2014)}]{ganatra2014few}
\bibinfo{author}{\bibfnamefont{R.}~\bibnamefont{Ganatra}} \bibnamefont{and}
  \bibinfo{author}{\bibfnamefont{Q.}~\bibnamefont{Zhang}},
  \bibinfo{journal}{ACS nano} \textbf{\bibinfo{volume}{8}},
  \bibinfo{pages}{4074} (\bibinfo{year}{2014}).

\bibitem[{\citenamefont{Geim and Grigorieva}(2013)}]{geim2013van}
\bibinfo{author}{\bibfnamefont{A.~K.} \bibnamefont{Geim}} \bibnamefont{and}
  \bibinfo{author}{\bibfnamefont{I.~V.} \bibnamefont{Grigorieva}},
  \bibinfo{journal}{Nature} \textbf{\bibinfo{volume}{499}},
  \bibinfo{pages}{419} (\bibinfo{year}{2013}).

\bibitem[{\citenamefont{Butler et~al.}(2013)\citenamefont{Butler, Hollen, Cao,
  Cui, Gupta, Guti{\'e}rrez, Heinz, Hong, Huang, Ismach
  et~al.}}]{butler2013progress}
\bibinfo{author}{\bibfnamefont{S.~Z.} \bibnamefont{Butler}},
  \bibinfo{author}{\bibfnamefont{S.~M.} \bibnamefont{Hollen}},
  \bibinfo{author}{\bibfnamefont{L.}~\bibnamefont{Cao}},
  \bibinfo{author}{\bibfnamefont{Y.}~\bibnamefont{Cui}},
  \bibinfo{author}{\bibfnamefont{J.~A.} \bibnamefont{Gupta}},
  \bibinfo{author}{\bibfnamefont{H.~R.} \bibnamefont{Guti{\'e}rrez}},
  \bibinfo{author}{\bibfnamefont{T.~F.} \bibnamefont{Heinz}},
  \bibinfo{author}{\bibfnamefont{S.~S.} \bibnamefont{Hong}},
  \bibinfo{author}{\bibfnamefont{J.}~\bibnamefont{Huang}},
  \bibinfo{author}{\bibfnamefont{A.~F.} \bibnamefont{Ismach}},
  \bibnamefont{et~al.}, \bibinfo{journal}{ACS nano}
  \textbf{\bibinfo{volume}{7}}, \bibinfo{pages}{2898} (\bibinfo{year}{2013}).

\bibitem[{\citenamefont{Bernardi et~al.}(2013)\citenamefont{Bernardi, Palummo,
  and Grossman}}]{bernardi2013extraordinary}
\bibinfo{author}{\bibfnamefont{M.}~\bibnamefont{Bernardi}},
  \bibinfo{author}{\bibfnamefont{M.}~\bibnamefont{Palummo}}, \bibnamefont{and}
  \bibinfo{author}{\bibfnamefont{J.~C.} \bibnamefont{Grossman}},
  \bibinfo{journal}{Nano letters} \textbf{\bibinfo{volume}{13}},
  \bibinfo{pages}{3664} (\bibinfo{year}{2013}).

\bibitem[{\citenamefont{Li and Zhu}(2015)}]{li2015two}
\bibinfo{author}{\bibfnamefont{X.}~\bibnamefont{Li}} \bibnamefont{and}
  \bibinfo{author}{\bibfnamefont{H.}~\bibnamefont{Zhu}},
  \bibinfo{journal}{Journal of Materiomics} \textbf{\bibinfo{volume}{1}},
  \bibinfo{pages}{33} (\bibinfo{year}{2015}).

\bibitem[{\citenamefont{Nayak et~al.}(2014)\citenamefont{Nayak, Bhattacharyya,
  Zhu, Liu, Wu, Pandey, Jin, Singh, Akinwande, and Lin}}]{nayak2014pressure}
\bibinfo{author}{\bibfnamefont{A.~P.} \bibnamefont{Nayak}},
  \bibinfo{author}{\bibfnamefont{S.}~\bibnamefont{Bhattacharyya}},
  \bibinfo{author}{\bibfnamefont{J.}~\bibnamefont{Zhu}},
  \bibinfo{author}{\bibfnamefont{J.}~\bibnamefont{Liu}},
  \bibinfo{author}{\bibfnamefont{X.}~\bibnamefont{Wu}},
  \bibinfo{author}{\bibfnamefont{T.}~\bibnamefont{Pandey}},
  \bibinfo{author}{\bibfnamefont{C.}~\bibnamefont{Jin}},
  \bibinfo{author}{\bibfnamefont{A.~K.} \bibnamefont{Singh}},
  \bibinfo{author}{\bibfnamefont{D.}~\bibnamefont{Akinwande}},
  \bibnamefont{and} \bibinfo{author}{\bibfnamefont{J.-F.} \bibnamefont{Lin}},
  \bibinfo{journal}{Nature communications} \textbf{\bibinfo{volume}{5}},
  \bibinfo{pages}{3731} (\bibinfo{year}{2014}).

\bibitem[{\citenamefont{Yue et~al.}(2012)\citenamefont{Yue, Kang, Shao, Zhang,
  Chang, Wang, Qin, and Li}}]{yue2012mechanical}
\bibinfo{author}{\bibfnamefont{Q.}~\bibnamefont{Yue}},
  \bibinfo{author}{\bibfnamefont{J.}~\bibnamefont{Kang}},
  \bibinfo{author}{\bibfnamefont{Z.}~\bibnamefont{Shao}},
  \bibinfo{author}{\bibfnamefont{X.}~\bibnamefont{Zhang}},
  \bibinfo{author}{\bibfnamefont{S.}~\bibnamefont{Chang}},
  \bibinfo{author}{\bibfnamefont{G.}~\bibnamefont{Wang}},
  \bibinfo{author}{\bibfnamefont{S.}~\bibnamefont{Qin}}, \bibnamefont{and}
  \bibinfo{author}{\bibfnamefont{J.}~\bibnamefont{Li}},
  \bibinfo{journal}{Physics Letters A} \textbf{\bibinfo{volume}{376}},
  \bibinfo{pages}{1166} (\bibinfo{year}{2012}).

\bibitem[{\citenamefont{Conley et~al.}(2013)\citenamefont{Conley, Wang,
  Ziegler, Haglund~Jr, Pantelides, and Bolotin}}]{conley2013bandgap}
\bibinfo{author}{\bibfnamefont{H.~J.} \bibnamefont{Conley}},
  \bibinfo{author}{\bibfnamefont{B.}~\bibnamefont{Wang}},
  \bibinfo{author}{\bibfnamefont{J.~I.} \bibnamefont{Ziegler}},
  \bibinfo{author}{\bibfnamefont{R.~F.} \bibnamefont{Haglund~Jr}},
  \bibinfo{author}{\bibfnamefont{S.~T.} \bibnamefont{Pantelides}},
  \bibnamefont{and} \bibinfo{author}{\bibfnamefont{K.~I.}
  \bibnamefont{Bolotin}}, \bibinfo{journal}{Nano letters}
  \textbf{\bibinfo{volume}{13}}, \bibinfo{pages}{3626} (\bibinfo{year}{2013}).

\bibitem[{\citenamefont{Pena-Alvarez et~al.}(2015)\citenamefont{Pena-Alvarez,
  del Corro, Morales-Garcia, Kavan, Kalbac, and Frank}}]{pena2015single}
\bibinfo{author}{\bibfnamefont{M.}~\bibnamefont{Pena-Alvarez}},
  \bibinfo{author}{\bibfnamefont{E.}~\bibnamefont{del Corro}},
  \bibinfo{author}{\bibfnamefont{A.}~\bibnamefont{Morales-Garcia}},
  \bibinfo{author}{\bibfnamefont{L.}~\bibnamefont{Kavan}},
  \bibinfo{author}{\bibfnamefont{M.}~\bibnamefont{Kalbac}}, \bibnamefont{and}
  \bibinfo{author}{\bibfnamefont{O.}~\bibnamefont{Frank}},
  \bibinfo{journal}{Nano letters} \textbf{\bibinfo{volume}{15}},
  \bibinfo{pages}{3139} (\bibinfo{year}{2015}).

\bibitem[{\citenamefont{Radisavljevic and
  Kis}(2013)}]{radisavljevic2013mobility}
\bibinfo{author}{\bibfnamefont{B.}~\bibnamefont{Radisavljevic}}
  \bibnamefont{and} \bibinfo{author}{\bibfnamefont{A.}~\bibnamefont{Kis}},
  \bibinfo{journal}{Nature materials} \textbf{\bibinfo{volume}{12}},
  \bibinfo{pages}{815} (\bibinfo{year}{2013}).

\bibitem[{\citenamefont{Ridolfi et~al.}(2015)\citenamefont{Ridolfi, Le, Rahman,
  Mucciolo, and Lewenkopf}}]{ridolfi2015tight}
\bibinfo{author}{\bibfnamefont{E.}~\bibnamefont{Ridolfi}},
  \bibinfo{author}{\bibfnamefont{D.}~\bibnamefont{Le}},
  \bibinfo{author}{\bibfnamefont{T.}~\bibnamefont{Rahman}},
  \bibinfo{author}{\bibfnamefont{E.}~\bibnamefont{Mucciolo}}, \bibnamefont{and}
  \bibinfo{author}{\bibfnamefont{C.}~\bibnamefont{Lewenkopf}},
  \bibinfo{journal}{Journal of Physics: Condensed Matter}
  \textbf{\bibinfo{volume}{27}}, \bibinfo{pages}{365501}
  (\bibinfo{year}{2015}).

\bibitem[{\citenamefont{Li et~al.}(2014{\natexlab{b}})\citenamefont{Li, Zhang,
  Guo, and Zhang}}]{li2014strain}
\bibinfo{author}{\bibfnamefont{W.}~\bibnamefont{Li}},
  \bibinfo{author}{\bibfnamefont{G.}~\bibnamefont{Zhang}},
  \bibinfo{author}{\bibfnamefont{M.}~\bibnamefont{Guo}}, \bibnamefont{and}
  \bibinfo{author}{\bibfnamefont{Y.-W.} \bibnamefont{Zhang}},
  \bibinfo{journal}{Nano Research} \textbf{\bibinfo{volume}{7}},
  \bibinfo{pages}{518} (\bibinfo{year}{2014}{\natexlab{b}}).

\bibitem[{\citenamefont{Yu et~al.}(2016)\citenamefont{Yu, Ong, Pan, Cui, Xin,
  Shi, Wang, Wu, Chen, Zhang et~al.}}]{yu2016realization}
\bibinfo{author}{\bibfnamefont{Z.}~\bibnamefont{Yu}},
  \bibinfo{author}{\bibfnamefont{Z.-Y.} \bibnamefont{Ong}},
  \bibinfo{author}{\bibfnamefont{Y.}~\bibnamefont{Pan}},
  \bibinfo{author}{\bibfnamefont{Y.}~\bibnamefont{Cui}},
  \bibinfo{author}{\bibfnamefont{R.}~\bibnamefont{Xin}},
  \bibinfo{author}{\bibfnamefont{Y.}~\bibnamefont{Shi}},
  \bibinfo{author}{\bibfnamefont{B.}~\bibnamefont{Wang}},
  \bibinfo{author}{\bibfnamefont{Y.}~\bibnamefont{Wu}},
  \bibinfo{author}{\bibfnamefont{T.}~\bibnamefont{Chen}},
  \bibinfo{author}{\bibfnamefont{Y.-W.} \bibnamefont{Zhang}},
  \bibnamefont{et~al.}, \bibinfo{journal}{Advanced Materials}
  \textbf{\bibinfo{volume}{28}}, \bibinfo{pages}{547} (\bibinfo{year}{2016}).

\bibitem[{\citenamefont{Xiao et~al.}(2014)\citenamefont{Xiao, Long, Li, Xu,
  Huang, and Gao}}]{xiao2014theoretical}
\bibinfo{author}{\bibfnamefont{J.}~\bibnamefont{Xiao}},
  \bibinfo{author}{\bibfnamefont{M.}~\bibnamefont{Long}},
  \bibinfo{author}{\bibfnamefont{X.}~\bibnamefont{Li}},
  \bibinfo{author}{\bibfnamefont{H.}~\bibnamefont{Xu}},
  \bibinfo{author}{\bibfnamefont{H.}~\bibnamefont{Huang}}, \bibnamefont{and}
  \bibinfo{author}{\bibfnamefont{Y.}~\bibnamefont{Gao}},
  \bibinfo{journal}{Scientific Reports} \textbf{\bibinfo{volume}{4}},
  \bibinfo{pages}{4327} (\bibinfo{year}{2014}).

\bibitem[{\citenamefont{Ma and Jena}(2014)}]{ma2014charge}
\bibinfo{author}{\bibfnamefont{N.}~\bibnamefont{Ma}} \bibnamefont{and}
  \bibinfo{author}{\bibfnamefont{D.}~\bibnamefont{Jena}},
  \bibinfo{journal}{Physical Review X} \textbf{\bibinfo{volume}{4}},
  \bibinfo{pages}{011043} (\bibinfo{year}{2014}).

\bibitem[{\citenamefont{Li et~al.}(2016{\natexlab{b}})\citenamefont{Li,
  Tsukagoshi, Orgiu, and Samor{\`\i}}}]{li2016charge}
\bibinfo{author}{\bibfnamefont{S.-L.} \bibnamefont{Li}},
  \bibinfo{author}{\bibfnamefont{K.}~\bibnamefont{Tsukagoshi}},
  \bibinfo{author}{\bibfnamefont{E.}~\bibnamefont{Orgiu}}, \bibnamefont{and}
  \bibinfo{author}{\bibfnamefont{P.}~\bibnamefont{Samor{\`\i}}},
  \bibinfo{journal}{Chemical Society Reviews} \textbf{\bibinfo{volume}{45}},
  \bibinfo{pages}{118} (\bibinfo{year}{2016}{\natexlab{b}}).

\bibitem[{\citenamefont{Ong and Fischetti}(2013)}]{ong2013mobility}
\bibinfo{author}{\bibfnamefont{Z.-Y.} \bibnamefont{Ong}} \bibnamefont{and}
  \bibinfo{author}{\bibfnamefont{M.~V.} \bibnamefont{Fischetti}},
  \bibinfo{journal}{Physical Review B} \textbf{\bibinfo{volume}{88}},
  \bibinfo{pages}{165316} (\bibinfo{year}{2013}).

\bibitem[{\citenamefont{Patil et~al.}(2017)\citenamefont{Patil, Sankeshwar, and
  Mulimani}}]{patil2017role}
\bibinfo{author}{\bibfnamefont{S.~B.} \bibnamefont{Patil}},
  \bibinfo{author}{\bibfnamefont{N.}~\bibnamefont{Sankeshwar}},
  \bibnamefont{and} \bibinfo{author}{\bibfnamefont{B.}~\bibnamefont{Mulimani}},
  \bibinfo{journal}{Journal of Physics: Condensed Matter}
  \textbf{\bibinfo{volume}{29}}, \bibinfo{pages}{485303}
  (\bibinfo{year}{2017}).

\bibitem[{\citenamefont{Daughton}(1999)}]{DAUGHTON1999334}
\bibinfo{author}{\bibfnamefont{J.}~\bibnamefont{Daughton}},
  \bibinfo{journal}{Journal of Magnetism and Magnetic Materials}
  \textbf{\bibinfo{volume}{192}}, \bibinfo{pages}{334} (\bibinfo{year}{1999}),
  ISSN \bibinfo{issn}{0304-8853},
  \urlprefix\url{https://www.sciencedirect.com/science/article/pii/S030488539800376X}.

\bibitem[{\citenamefont{Lenz}(1990)}]{56910}
\bibinfo{author}{\bibfnamefont{J.}~\bibnamefont{Lenz}},
  \bibinfo{journal}{Proceedings of the IEEE} \textbf{\bibinfo{volume}{78}},
  \bibinfo{pages}{973} (\bibinfo{year}{1990}).

\bibitem[{\citenamefont{Rife et~al.}(2003)\citenamefont{Rife, Miller, Sheehan,
  Tamanaha, Tondra, and Whitman}}]{RIFE2003209}
\bibinfo{author}{\bibfnamefont{J.}~\bibnamefont{Rife}},
  \bibinfo{author}{\bibfnamefont{M.}~\bibnamefont{Miller}},
  \bibinfo{author}{\bibfnamefont{P.}~\bibnamefont{Sheehan}},
  \bibinfo{author}{\bibfnamefont{C.}~\bibnamefont{Tamanaha}},
  \bibinfo{author}{\bibfnamefont{M.}~\bibnamefont{Tondra}}, \bibnamefont{and}
  \bibinfo{author}{\bibfnamefont{L.}~\bibnamefont{Whitman}},
  \bibinfo{journal}{Sensors and Actuators A: Physical}
  \textbf{\bibinfo{volume}{107}}, \bibinfo{pages}{209} (\bibinfo{year}{2003}).

\bibitem[{\citenamefont{Baselt et~al.}(1998)\citenamefont{Baselt, Lee, Natesan,
  Metzger, Sheehan, and Colton}}]{baselt1998biosensor}
\bibinfo{author}{\bibfnamefont{D.~R.} \bibnamefont{Baselt}},
  \bibinfo{author}{\bibfnamefont{G.~U.} \bibnamefont{Lee}},
  \bibinfo{author}{\bibfnamefont{M.}~\bibnamefont{Natesan}},
  \bibinfo{author}{\bibfnamefont{S.~W.} \bibnamefont{Metzger}},
  \bibinfo{author}{\bibfnamefont{P.~E.} \bibnamefont{Sheehan}},
  \bibnamefont{and} \bibinfo{author}{\bibfnamefont{R.~J.}
  \bibnamefont{Colton}}, \bibinfo{journal}{Biosensors and Bioelectronics}
  \textbf{\bibinfo{volume}{13}}, \bibinfo{pages}{731} (\bibinfo{year}{1998}).

\bibitem[{\citenamefont{Lundstrom}(2002)}]{lundstrom2002fundamentals}
\bibinfo{author}{\bibfnamefont{M.}~\bibnamefont{Lundstrom}},
  \bibinfo{journal}{Measurement Science and Technology}
  \textbf{\bibinfo{volume}{13}}, \bibinfo{pages}{230} (\bibinfo{year}{2002}).

\bibitem[{\citenamefont{Mandia et~al.}(2021)\citenamefont{Mandia, Muralidharan,
  Choi, Lee, and Bhattacharjee}}]{mandia2021ammcr}
\bibinfo{author}{\bibfnamefont{A.~K.} \bibnamefont{Mandia}},
  \bibinfo{author}{\bibfnamefont{B.}~\bibnamefont{Muralidharan}},
  \bibinfo{author}{\bibfnamefont{J.-H.} \bibnamefont{Choi}},
  \bibinfo{author}{\bibfnamefont{S.-C.} \bibnamefont{Lee}}, \bibnamefont{and}
  \bibinfo{author}{\bibfnamefont{S.}~\bibnamefont{Bhattacharjee}},
  \bibinfo{journal}{Computer Physics Communications}
  \textbf{\bibinfo{volume}{259}}, \bibinfo{pages}{107697}
  (\bibinfo{year}{2021}).

\bibitem[{\citenamefont{Singh}(2007)}]{singh2007electronic}
\bibinfo{author}{\bibfnamefont{J.}~\bibnamefont{Singh}},
  \emph{\bibinfo{title}{Electronic and optoelectronic properties of
  semiconductor structures}} (\bibinfo{publisher}{Cambridge University Press},
  \bibinfo{year}{2007}).

\bibitem[{\citenamefont{Ferry}(2016)}]{ferry2016semiconductor}
\bibinfo{author}{\bibfnamefont{D.~K.} \bibnamefont{Ferry}},
  \emph{\bibinfo{title}{Semiconductor transport}} (\bibinfo{publisher}{CRC
  Press}, \bibinfo{year}{2016}).

\bibitem[{\citenamefont{Mandia et~al.}(2023)\citenamefont{Mandia, Kumar, Koshi,
  Lee, Bhattacharjee, and Muralidharan}}]{mandia2023high}
\bibinfo{author}{\bibfnamefont{A.~K.} \bibnamefont{Mandia}},
  \bibinfo{author}{\bibfnamefont{R.}~\bibnamefont{Kumar}},
  \bibinfo{author}{\bibfnamefont{N.~A.} \bibnamefont{Koshi}},
  \bibinfo{author}{\bibfnamefont{S.-C.} \bibnamefont{Lee}},
  \bibinfo{author}{\bibfnamefont{S.}~\bibnamefont{Bhattacharjee}},
  \bibnamefont{and}
  \bibinfo{author}{\bibfnamefont{B.}~\bibnamefont{Muralidharan}},
  \bibinfo{journal}{arXiv preprint arXiv:2301.11676}  (\bibinfo{year}{2023}).

\bibitem[{\citenamefont{Rode}(1973)}]{rode1973theory}
\bibinfo{author}{\bibfnamefont{D.}~\bibnamefont{Rode}},
  \bibinfo{journal}{physica status solidi (b)} \textbf{\bibinfo{volume}{55}},
  \bibinfo{pages}{687} (\bibinfo{year}{1973}).

\bibitem[{\citenamefont{Rode et~al.}(1983)\citenamefont{Rode, Wolfe, and
  Stillman}}]{rode1983magnetic}
\bibinfo{author}{\bibfnamefont{D.}~\bibnamefont{Rode}},
  \bibinfo{author}{\bibfnamefont{C.}~\bibnamefont{Wolfe}}, \bibnamefont{and}
  \bibinfo{author}{\bibfnamefont{G.}~\bibnamefont{Stillman}},
  \bibinfo{journal}{Journal of applied physics} \textbf{\bibinfo{volume}{54}},
  \bibinfo{pages}{10} (\bibinfo{year}{1983}).

\bibitem[{\citenamefont{Kumar et~al.}(2023)\citenamefont{Kumar, Mandia, Singh,
  and Muralidharan}}]{kumar2023advancing}
\bibinfo{author}{\bibfnamefont{R.}~\bibnamefont{Kumar}},
  \bibinfo{author}{\bibfnamefont{A.~K.} \bibnamefont{Mandia}},
  \bibinfo{author}{\bibfnamefont{A.}~\bibnamefont{Singh}}, \bibnamefont{and}
  \bibinfo{author}{\bibfnamefont{B.}~\bibnamefont{Muralidharan}},
  \bibinfo{journal}{Physical Review B} \textbf{\bibinfo{volume}{107}},
  \bibinfo{pages}{235303} (\bibinfo{year}{2023}).

\bibitem[{\citenamefont{Zou et~al.}(2010)\citenamefont{Zou, Hong, Keefer, and
  Zhu}}]{zou2010deposition}
\bibinfo{author}{\bibfnamefont{K.}~\bibnamefont{Zou}},
  \bibinfo{author}{\bibfnamefont{X.}~\bibnamefont{Hong}},
  \bibinfo{author}{\bibfnamefont{D.}~\bibnamefont{Keefer}}, \bibnamefont{and}
  \bibinfo{author}{\bibfnamefont{J.}~\bibnamefont{Zhu}},
  \bibinfo{journal}{Physical review letters} \textbf{\bibinfo{volume}{105}},
  \bibinfo{pages}{126601} (\bibinfo{year}{2010}).

\bibitem[{\citenamefont{Nag and Dutta}(1975)}]{nag1975galvanomagnetic}
\bibinfo{author}{\bibfnamefont{B.}~\bibnamefont{Nag}} \bibnamefont{and}
  \bibinfo{author}{\bibfnamefont{G.}~\bibnamefont{Dutta}},
  \bibinfo{journal}{physica status solidi (b)} \textbf{\bibinfo{volume}{71}},
  \bibinfo{pages}{401} (\bibinfo{year}{1975}).

\bibitem[{\citenamefont{Rode}(1975)}]{rode1975low}
\bibinfo{author}{\bibfnamefont{D.}~\bibnamefont{Rode}}, in
  \emph{\bibinfo{booktitle}{Semiconductors and semimetals}}
  (\bibinfo{publisher}{Elsevier}, \bibinfo{year}{1975}),
  vol.~\bibinfo{volume}{10}, pp. \bibinfo{pages}{1--89}.

\bibitem[{\citenamefont{Rode and Knight}(1971)}]{rode1971electron}
\bibinfo{author}{\bibfnamefont{D.}~\bibnamefont{Rode}} \bibnamefont{and}
  \bibinfo{author}{\bibfnamefont{S.}~\bibnamefont{Knight}},
  \bibinfo{journal}{Physical Review B} \textbf{\bibinfo{volume}{3}},
  \bibinfo{pages}{2534} (\bibinfo{year}{1971}).

\end{thebibliography}
\end{document}